\documentclass[12pt]{iopart}

\usepackage{amsfonts,graphicx,soul,mathrsfs,bm,xcolor,hyperref}
\usepackage{amsmath} 
\usepackage[square,numbers,sort&compress]{natbib}
\usepackage{upgreek}
\usepackage{algorithm}
\usepackage[noend]{algpseudocode}
\usepackage{relsize}
\usepackage{xcolor}

\begin{document}
\vbox{}
\vspace{-14mm}

\title[\textsc{Multipoint-BAX:} Tuning Particle Accelerator Emittance via Virtual Objectives]{\textsc{Multipoint-BAX}: A New Approach for Efficiently Tuning Particle Accelerator Emittance via Virtual Objectives}

\vspace{3mm}
\author{\footnotesize Sara Ayoub Miskovich$^{1\dagger}$, Willie Neiswanger$^{2\dagger}$, William Colocho$^1$, Claudio Emma$^1$, Jacqueline Garrahan$^1$, Timothy Maxwell$^1$, Christopher Mayes$^1$, Stefano Ermon$^2$, Auralee Edelen$^1$, Daniel Ratner$^1$}
\vspace{4mm}
\address{$^1$SLAC National Accelerator Laboratory, Menlo Park, California, USA\\
$^2$Department of Computer Science, Stanford University, Stanford, California, USA}
\footnotetext{\scriptsize Shared co-first author.\\E-mail: \textit{smiskov@slac.stanford.edu}, \textit{neiswanger@cs.stanford.edu}, \textit{edelen@slac.stanford.edu}, \textit{dratner@slac.stanford.edu}}

\vspace{4mm}
\begin{abstract}
Although beam emittance is critical for the performance of high-brightness accelerators, optimization is often time limited as emittance calculations, commonly done via quadrupole scans, are typically slow. Such calculations are a type of \emph{multipoint query}, i.e. each query requires multiple secondary measurements. Traditional black-box optimizers such as Bayesian optimization are slow and inefficient when dealing with such objectives as they must acquire the full series of measurements, but return only the emittance, with each query. We propose a new information-theoretic algorithm, \textsc{Multipoint-BAX}, for black-box optimization on multipoint queries, which queries and models individual beam-size measurements using techniques from Bayesian Algorithm Execution (BAX). Our method avoids the slow multipoint query on the accelerator by acquiring points through a \emph{virtual objective}, i.e. calculating the emittance objective from a fast learned model rather than directly from the accelerator. We use \textsc{Multipoint-BAX} to minimize emittance at the Linac Coherent Light Source (LCLS) and the Facility for Advanced Accelerator Experimental Tests II (FACET-II). In simulation, our method is 20$\times$ faster and more robust to noise compared to existing methods. In live tests, it matched the hand-tuned emittance at FACET-II and achieved a 24\% lower emittance than hand-tuning at LCLS.
Our method represents a conceptual shift for optimizing multipoint queries, and we anticipate that it can be readily adapted to similar problems in particle accelerators and other scientific instruments.
\end{abstract}

\vspace{-5mm}
\section{Introduction}
\label{sec:intro}
\vspace{-1mm}

Black-box optimizers \cite{jones1998efficient, brochu2010}, including the Nelder-Mead simplex method, coordinate descent, and genetic algorithms, are used widely in online optimization of particle accelerators \cite{rcds_demo, rcds, ES, HuangRSimplex, ScheinkerSP32018, BerganGA2019, ES2019} as well as other areas of science \cite{terayama2021black, char2019offline, ueno2016combo}.
Bayesian optimization (BO) \cite{kushner1963new, mockus1975} is an especially appealing approach for use in black-box systems in which each function query is expensive (in time or other costs),
a common situation for scientific and industrial instruments. BO has been used extensively for optimization of scientific instruments, including particle accelerators \cite{McIntire:2016fnl,kirschner2019,duris2020,Shalloo_2020,Roussel_2021}.

For applications where the optimization objective is not a direct system observable, but rather is computed from a series of secondary measurements, even BO may be too inefficient. For example, a single query might require a scan of another control variable, a series of expensive computations, or an internal optimization process. We refer to this type of query as a \emph{multipoint query}. Black-box optimization tasks with multipoint queries are encountered in a variety of fields, such as material science \cite{yamashita}, nuclear astrophysics \cite{miskovich}, aerospace engineering \cite{liem2015}, bioinformatics \cite{terayama}, in addition to particle accelerators \cite{PhysRevAccelBeams.23.114201}. When dealing with such tasks, traditional optimization methods, including BO, exhibit two types of inefficiencies. First, each query requires a set number of secondary measurement points to calculate the objective, even if a subset of those points already reveals the system settings to be suboptimal. Second, because the black-box function returns only the objective to the optimization algorithm, the full information acquired within the black-box function evaluation (e.g. the full series of secondary measurements) is not shared across different optimization steps, and this information loss leads to inefficient sampling. The result is that each multipoint query is both \emph{expensive and information-poor}. 
When time is a limiting factor, optimizer inefficiency may reduce the number of control variables that can be optimized, curtail the ability to find a global optimum, or make the optimization task too costly to even attempt in practice. Optimizer inefficiency is particularly problematic for scientific instruments that are in high demand.

The motivating example for this paper is beam-emittance minimization in particle accelerators. The emittance of a charged particle beam, defined as the volume of particles in position-momentum phase space \cite{minty}, is an important parameter for a variety of accelerator applications. For accelerator-based light sources, emittance determines the X-ray beam brightness, limiting the shortest wavelength available at X-ray free-electron lasers (XFELs) \cite{Emma2010} and affecting the output X-ray power by orders of magnitude \cite{kim2007}. Emittance is especially important for the next generation of undulator designs.
For example, at the LCLS-II high-energy upgrade (LCLS-II-HE) \cite{osti_1634206}, a modest emittance reduction from 0.6 mm$\cdot$mrad to 0.4 mm$\cdot$mrad is projected to increase the maximum photon energy that can reach saturation from 7.3 keV to 10.5 keV, enabling atomic resolution of biomolecules (see \ref{appendixlclsii} for details). For colliders, small emittance is critical to maximize luminosity \cite{PhysRevSTAB.4.053501,PhysRevSTAB.18.101002}. Beams with high emittance ratios are needed for a variety of new light source and collider designs \cite{PhysRevSTAB.9.031001}, and for injection into components of compact accelerators, such as dielectric wakefield accelerating cavities \cite{ODY201775}.

The accelerator settings need to be periodically re-optimized, or \emph{tuned}, to achieve acceptable emittance values. Emittance depends on the initial beam conditions at the particle source (e.g. the photocathode and associated drive laser for electron accelerators), along with the combined effects of adjustable accelerator settings. The initial conditions can change as a result of both intended beam property adjustments (e.g. changes to the beam charge) and unintended drift over time (e.g. uncontrolled time-varying changes to the drive laser output). Given the importance of the beam emittance to accelerator performance, emittance optimization would ideally be a routine task for accelerator operations.

However, one of the most common methods for determining the transverse emittance involves scanning a focusing magnet's strength while observing the resulting change in beam size \cite{minty}, i.e. the emittance is calculated from a multipoint query. Each step of the optimizer thus involves choosing a configuration in the control domain (the settings of the accelerator), then taking a set of measurements over a defined range in the secondary domain (in this case, beam-size measurements at different magnet settings), and finally fitting the measured points in the secondary domain to calculate an emittance value to the optimization algorithm.
The inefficiency of measuring emittance severely limits the number of tuning opportunities at most accelerator facilities, despite the importance of achieving small emittance. For example, at the  Linac Coherent Light Source (LCLS) \cite{Emma2010}, injector emittance is typically tuned only after returning from a machine shut-down or a substantial change to the target beam parameters (such as the charge), and it is done almost entirely manually by human experts. There is a clear need for more efficient algorithms for emittance optimization.

In this paper, we develop a new information-theoretic algorithm, \textsc{Multipoint-BAX}, as a sample-efficient method for optimization tasks with multipoint queries, using techniques adapted from Bayesian Algorithm Execution (BAX) \cite{neiswanger2021bayesian}.
In contrast to BO, which directly models the objective function, \textsc{Multipoint-BAX} builds a system surrogate model in the joint control-measurement domain, defined by the Cartesian product of the optimization’s control and secondary domain. \textsc{Multipoint-BAX} then guides the acquisition with a \emph{virtual} objective, i.e. one that is calculated on the surrogate model rather than obtained directly from the machine. By modeling the \textit{joint control-measurement domain}, as opposed to modeling only the \textit{objective} of a multipoint query, our procedure maximizes the information gained from individual measurements. In addition, by employing a virtual objective, \textsc{Multipoint-BAX} can query individual points in the joint control-measurement domain, avoiding the need for multiple measurements at each step. We emphasize that querying and modeling direct observables (e.g. beam sizes using \textsc{Multipoint-BAX}), rather than the objective (e.g. emittance using BO) represents a significant conceptual shift in black-box optimization with multipoint queries. 
 
In summary, this paper introduces a new information-theoretic algorithm, \textsc{Multipoint-BAX}, for optimization tasks with multipoint queries. We present its application to the challenging and outstanding problem of emittance tuning in accelerators, with substantial improvement in sample-efficiency. First, we present the challenges of emittance optimization to motivate this work. Second, we describe how \textsc{Multipoint-BAX} can improve efficiency for general optimization tasks involving multipoint queries. Third, for the specific case of emittance optimization, we show our method to be sample-efficient compared to both BO and the Nelder-Mead simplex algorithm \cite{nelder-mead}, using a simulation environment of the LCLS injector. Finally, we show experimental results for \textsc{Multipoint-BAX} in minimization of electron-beam emittance at both LCLS and the Facility for Advanced Accelerator Experimental Tests II (FACET-II) \cite{yakimenko2018} at SLAC National Accelerator Laboratory (SLAC), with the former achieving 24\% lower emittance than was found by manual tuning. We anticipate this approach to be widely adaptable to optimization problems involving multipoint queries on real-world instruments in science and engineering.

\section{The Challenges of Emittance Optimization}

This work is motivated by the challenges typically faced when optimizing the emittance in particle accelerators. Despite the importance of emittance to the lasing process, optimization is not a part of standard daily operations. Tuning time is limited by the need to minimize beam interruptions to users, which roughly translates into a limit on the number of queries to the system available for each tuning event. Emittance tuning is particularly slow because the exact response of the emittance to accelerator settings is unknown, and thus emittance is typically tuned as a black-box: iteratively adjusting the injector settings and observing the outcome. Using the most common method of emittance measurement, called the ``quadrupole scan'' method, a series of beam-size measurements is needed to calculate the emittance at each configuration of settings, making the query at each iteration particularly slow \cite{minty}. As a result, emittance tuning is typically limited to recovery from scheduled down-times and when switching to non-standard beam setups. Note that while single-shot emittance measurements for accelerators do exist (e.g. multi-slit and pepper-pot masks \cite{WANG1991190,2012AIPC.1507..757T,Zhang:1996af,Strehl2006}), they are generally not suitable for all types of beams and beamline layouts, including beams at the LCLS. 

With standard black-box emittance optimization using the quadrupole scan method, each step involves choosing a new control variable configuration (one or more accelerator injector magnet settings), running a subroutine that measures the beam size at multiple settings of a downstream quadrupole magnet, and calculating the emittance from the set of beam sizes observed. At LCLS, the subroutine is very time-consuming. When factoring in the need for robustness in automated tuning, we found that in practice we need to acquire an average of 18 beam-size measurements per emittance calculation, equivalent to at least 5 minutes of beam time per query using wire scanners.

\begin{figure}
\centering
\includegraphics[width=0.7\linewidth]{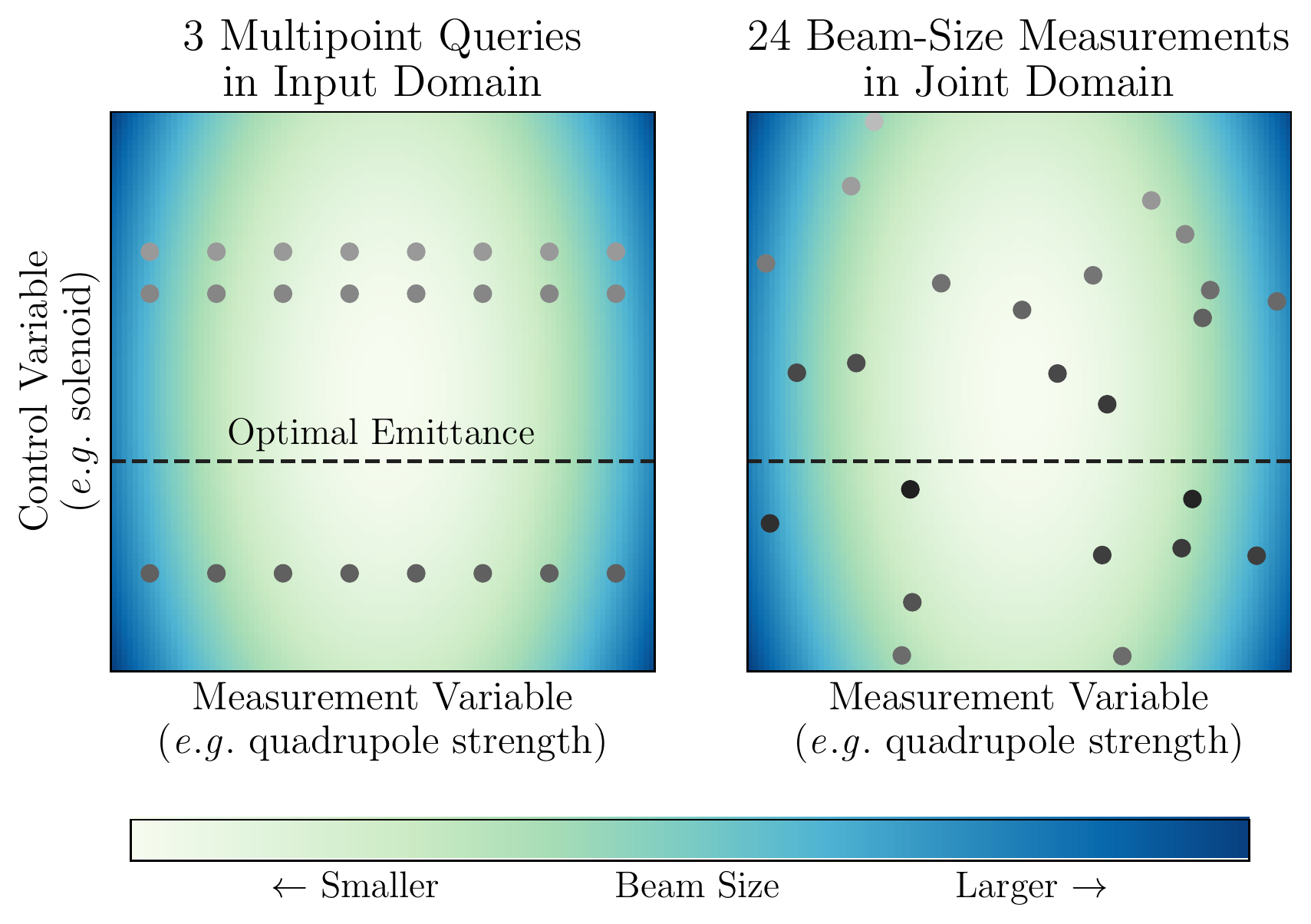}
\caption{Querying the joint control-measurement domain. \textmd{Left: Illustration of multipoint queries when tuning emittance with a standard black-box optimizer. The background color denotes the measured quantity (beam size), while the point marker color denotes the objective value (emittance) of the resulting multipoint query. The dashed line shows the optimal setting of the control variable. Optimizing the beam size is not equivalent to optimizing the emittance objective.
Right: Illustration of the same number of beam-size measurements when optimizing via queries in the joint domain to maximize information gain of each measurement.}}
\label{fig:sampling} 
\end{figure}

Figure~\ref{fig:sampling} (Left) highlights several problems encountered using BO---or other standard model-based optimizers---to tune emittance. As an illustration, we consider a typical emittance tuning task involving an eight-point calculation. First, BO must execute a full eight (or more) measurement scan at each query, even if the first beam size indicates the control setting is poor. Second, despite the wealth of information in the beam-size measurements, BO only uses the calculated emittance to build its model. As a result, different steps of the optimizer acquire nearly redundant beam-size measurements. Third, the measurement variable settings needed for an accurate emittance calculation depend on the control variables, which sometimes lead to scans that are unsuitable for fitting emittance. In BO, this valuable information obtained during a failed scan would be discarded, despite being helpful for characterizing the beam-size response to the control variable. 

\section{\textsc{Multipoint-BAX} for Efficient Emittance Optimization}

To overcome the challenges described above and improve the efficiency of optimization with multipoint queries, we propose to model the joint control-measurement domain, i.e. to learn a model of the beam size as a function of both the control and measurement variables. Figure \ref{fig:sampling} (Right) presents an illustration of how joint optimization samples individual beam-size measurements to maximize the information gained with each measurement. 
We highlight that joint optimization requires minimization of a computed function (e.g. emittance) operating on a learned model of a direct observable (e.g. beam size). However, existing standard model-based optimization methods such as BO are typically restricted to modeling and optimizing the same quantity. In the case of emittance tuning, standard BO can only model and optimize the computed emittance from a full quadrupole scan at each query point in the control domain. 
Here we introduce a novel method which solves all three of the above mentioned problems, sampling individual points from the joint domain, modeling all acquired information, and separating the modeling space from the optimization objective.

\begin{figure}
\begin{center}
\centering
\includegraphics[width=0.7\linewidth]{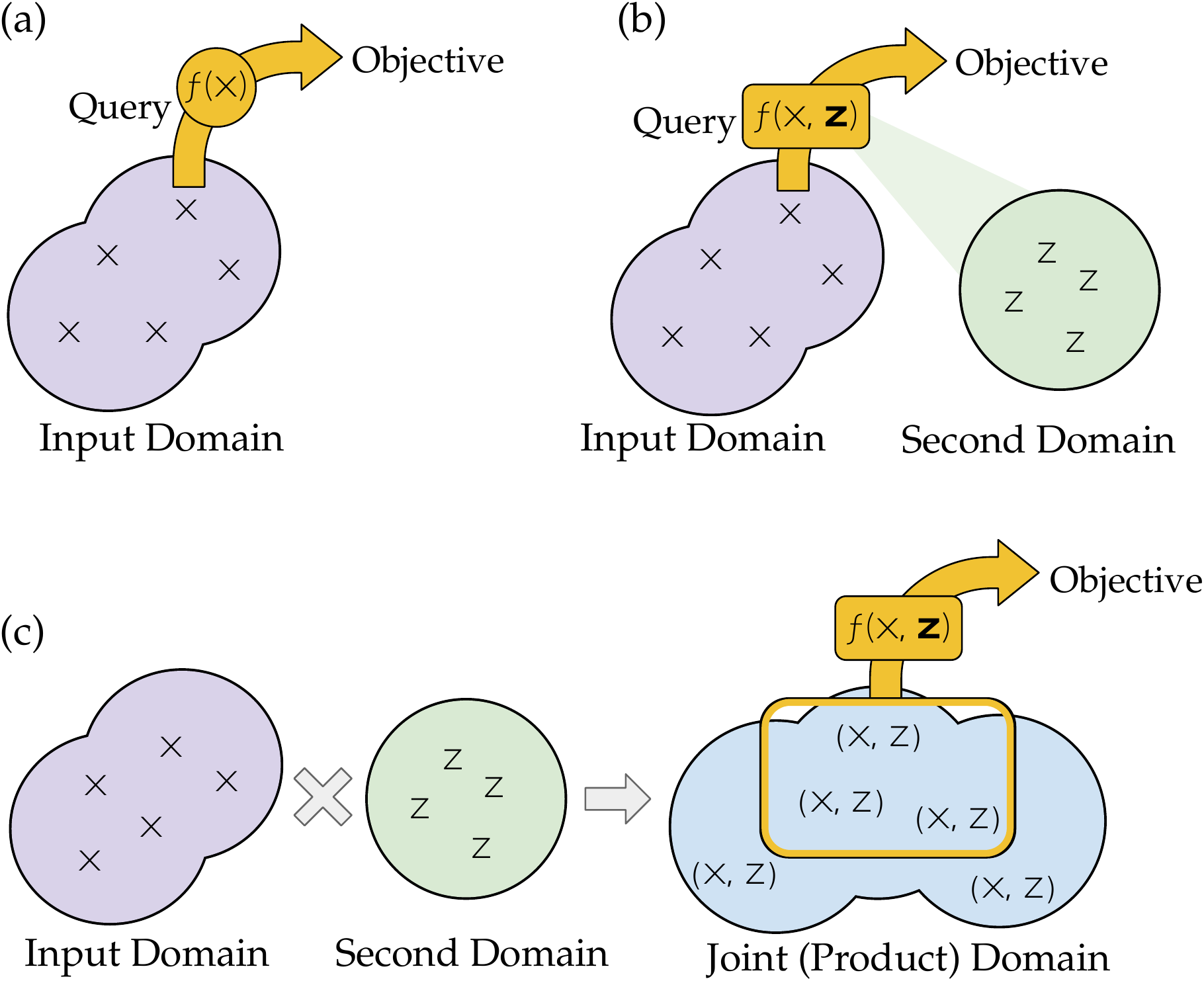}
\end{center}
\vspace{-3mm}
\caption{Optimization in the joint control-measurement domain. \textmd{(a) A query of objective function $f(\text{x})$ is made at control setting $\text{x}$ in the control domain. (b) A multipoint query at a control setting $\text{x}$ requires a vector of measurements at a set of points $\textbf{z}$ in a second domain. (c) The multipoint scenario is converted to an adaptive sampling problem in joint control-measurement domain.}}
\label{fig:domain}
\vspace{-2mm}
\end{figure}

We develop an information-based method, \textsc{Multipoint-BAX}, that allows us to model a direct observable that is \emph{different} than the objective. 
In the case of emittance tuning, we query a single point in the joint control-measurement domain, thereby observing a single beam size, rather than performing a full scan of beam size measurements to calculate the emittance objective. \textsc{Multipoint-BAX} then builds a model of the physical observable (beam size) with respect to both the control variables and the measurement variable.
However, as we do not wish to optimize the beam size, we introduce a \emph{virtual emittance objective} calculated on our learned model of beam sizes. This virtual objective is obtained by executing the full emittance calculation on function samples drawn from the learned model's posterior distribution; i.e. the emittance is calculated on the virtual beam size model, not on direct measurements. The next acquisition point is then chosen using an information-based strategy, which involves maximizing the mutual information between the model's posterior predictive distribution and the minimizer of the virtual objective.

\begin{figure}
\centering
\includegraphics[width=0.75\linewidth]{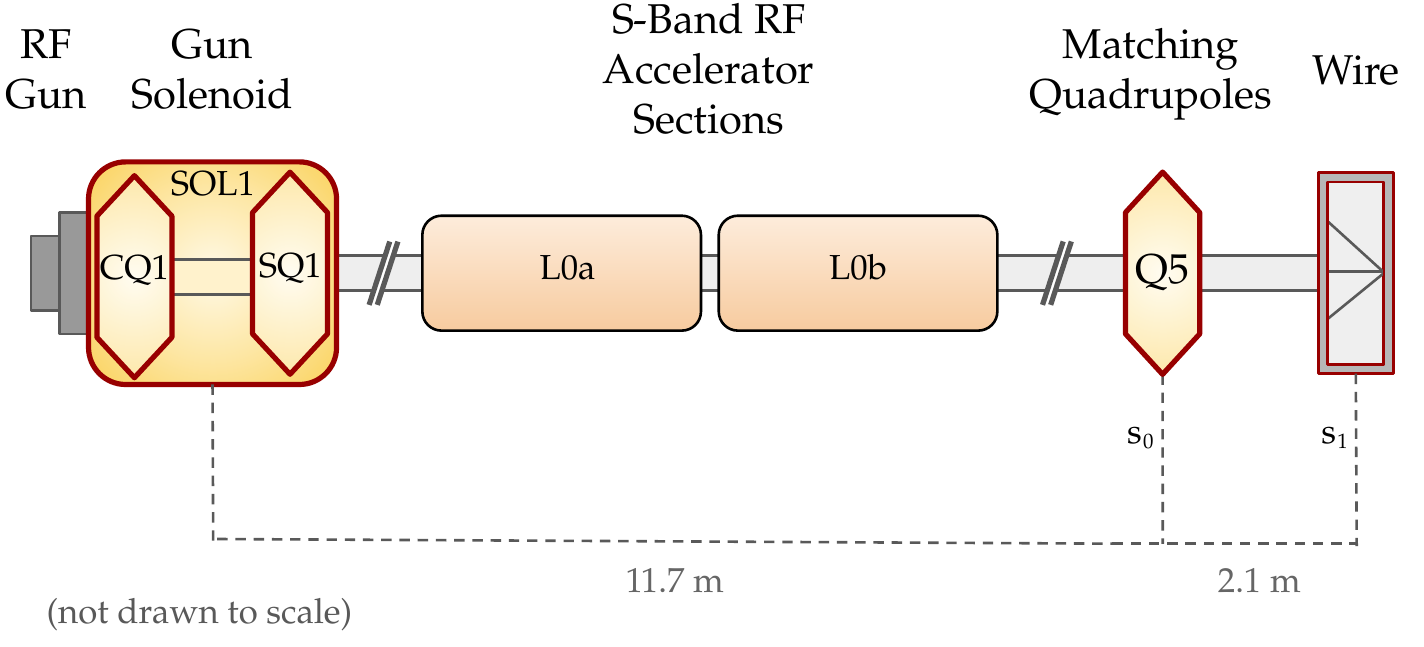}
\vspace{-2mm}
\caption{LCLS photoinjector layout. \textmd{The solenoid (SOL1) and corrector quadrupole (CQ1 and SQ1) settings comprise the control variables in our emittance optimization setting, and the quadrupole (Q5) setting is the measurement variable. The wire is used for beam-size measurements.}}
\label{fig:layout}
\vspace{-3mm}
\end{figure}

In the LCLS emittance application, \textsc{Multipoint-BAX} models the beam size with respect to several injector magnet settings. The three primary emittance-tuning variables at LCLS are the solenoid magnet (SOL1) and two corrector quadrupoles (CQ1 and SQ1) \cite{lclsinj}. Figure \ref{fig:layout} shows a simplified layout of the LCLS injector and emittance calculation setup. To run an emittance calculation for an injector configuration, a ``quadrupole scan'' \cite{minty} changes a quadrupole at location $s_0$ (denoted Q5), while measuring the beam size on a wire at location $s_1$.  We reconstruct the beam matrix elements by solving the system of equations
\begin{equation}
    \mathbf{\Sigma}_k^1 = \mathbf{R}_k\cdot\mathbf{\Sigma}^0 \cdot \mathbf{R}_k^\top,
    \label{eq:sigma}
\end{equation}
where $\mathbf{R}_k$ is the total transfer matrix (which describes the change in beam shape from $s_0$ to $s_1$) for the $k$th quadrupole setting, $\mathbf{\Sigma}^0$ is the beam matrix at location $s_0$, and $\mathbf{\Sigma}_k^1$ is the beam matrix at location $s_1$ for the $k$th quadrupole setting. Each 2$\times$2 beam matrix $\mathbf{\Sigma}$ defines the beam phase space through four elements: $\sigma_{11}$, $\sigma_{12} = \sigma_{21}$, and $\sigma_{22}$.
Each measurement at quadrupole setting $k$ returns the $\sigma_{11}$ element of $\mathbf{\Sigma}_k^1$; with sufficient measurements, we can use Eq~\eqref{eq:sigma} to solve for all three unknown elements of $\mathbf{\Sigma}^0$. We then obtain the single plane emittance from the $\mathbf{\Sigma}^0$ matrix using the relationship
\begin{equation}
    \epsilon = \sqrt{\sigma_{11}\sigma_{22}-\sigma_{12}^2}.
    \label{eq:emit}
\end{equation}
Equation \eqref{eq:emit} is computed for both transverse planes, $X$ and $Y$, and the normalized geometric mean gives the final emittance. For more details, see \cite{miskovich2022}. By applying this procedure to predictions from a beam-size model (rather than beam-size measurements), Equation~\eqref{eq:emit} produces a \emph{virtual} emittance objective.

\begin{figure}
\centering
\includegraphics[width=0.75\linewidth]{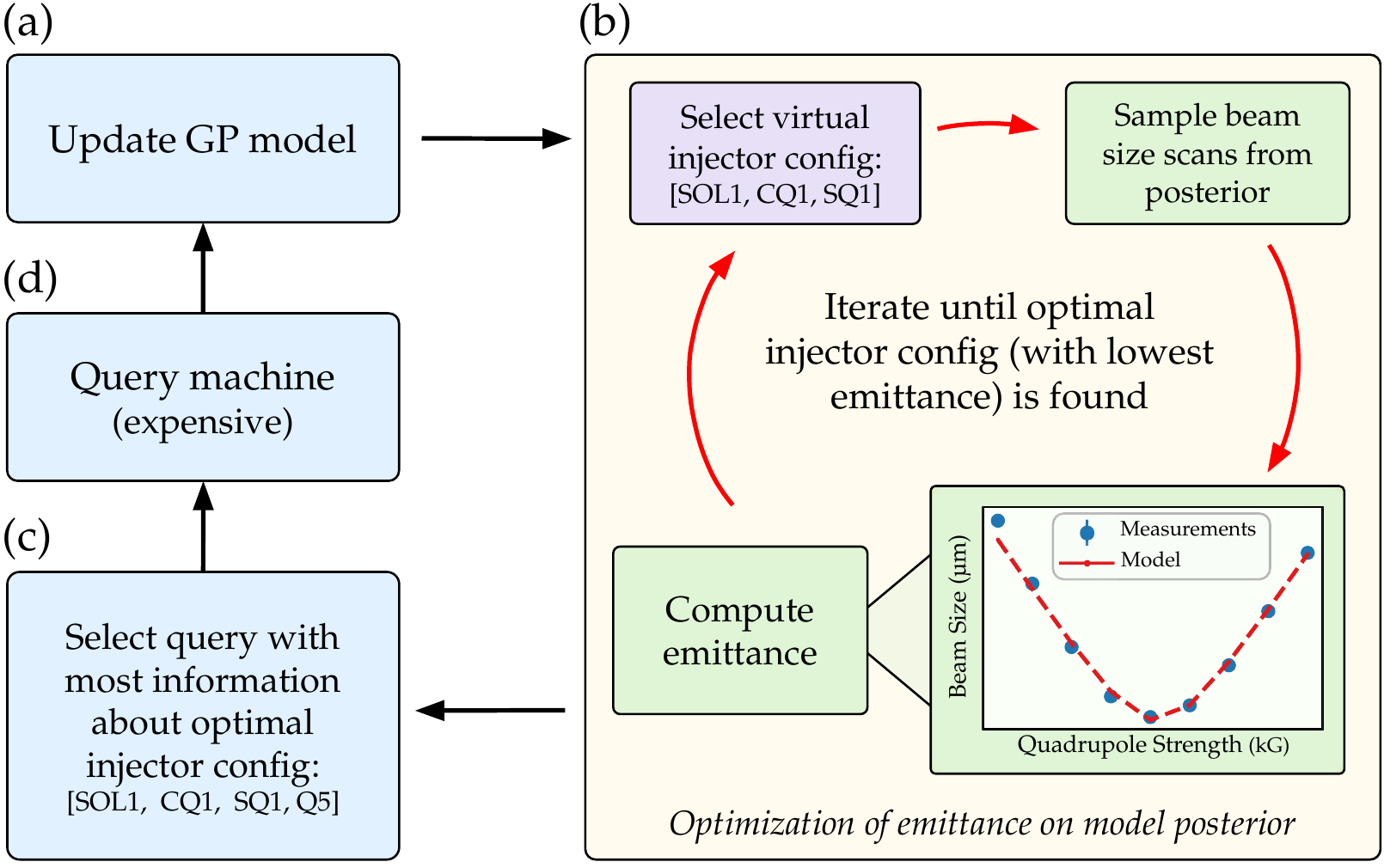}
\caption{\label{fig:diagram} Diagram of the \textsc{Multipoint-BAX} procedure for emittance optimization. \textmd{(a) A GP models the beam-size function with respect to the three control (SOL1, CQ1, and SQ1) and one measurement (Q5) variable, based on experimental data. (b) The emittance value and uncertainty of different injector configurations are assessed using cheap queries of the GP posterior. (c) A new experimental data point to query is suggested. (d) The machine is queried for a beam size. The colors of the boxes in this diagram match the colors of the domains illustrated in Figure~\ref{fig:domain}.}}
\vspace{-3mm}
\end{figure}

The full \textsc{Multipoint-BAX} emittance optimization procedure is illustrated in Figure~\ref{fig:diagram}. The procedure is initialized with randomly sampled points from the joint control-measurement domain. Then, at each iteration, the algorithm queries the beam size at a single value of the Q5, SOL1, CQ1 and SQ1 variables, constructing a model of the beam size as a function of both control (Q5) and measurement variables (SOL1, CQ1, and SQ1). This internal model represents the prior distribution on the function describing the beam-size response to the quadrupole strength, and is implemented in this work by a Gaussian process (GP) model; we parameterize the GP with a radial basis function kernel (for all input dimensions) with a fixed likelihood noise scale estimated offline from prior beam size measurements, where we fit kernel hyperparameters (lengthscales and output scale) by maximizing the marginal log likelihood \cite{gardner2018gpytorch, kandasamy2020tuning}.

During the internal step (Figure~\ref{fig:diagram}b), \textsc{Multipoint-BAX} executes the emittance calculation on a predefined number of quadrupole scans drawn from the model's posterior distribution, with each \emph{virtual} scan producing a \emph{virtual} objective. Finally, the information-based acquisition function selects a new beam-size measurement to query on the accelerator at the injector configuration that maximizes the mutual information between the accelerator settings and the virtual objective. Note that the query that maximizes the information gain is not necessarily the configuration with the lowest emittance based on the posterior samples. After measuring a new beam size, the algorithm updates the GP model and begins a new iteration. We emphasize that in a \textsc{Multipoint-BAX} emittance optimization, emittance calculations are only executed \textit{virtually on function samples drawn from the posterior distribution of the learned GP model, and never on the machine itself}.

In general, any automated emittance tuning approach requires a robust emittance calculation procedure. While \textsc{Multipoint-BAX} never executes a full beam-size scan for an emittance calculation on the real machine, an equivalent method executes on the model posterior. In this paper we used the newly-developed \verb|pyemittance| package \cite{miskovich2022} that provides an adaptive emittance procedure. More details regarding the emittance calculation are described in \ref{appendixemitmeas}.

In Algorithm~\ref{alg:multipointbax}, we provide pseudocode for our \textsc{Multipoint-BAX} procedure for emittance tuning.
Here, we use $\tilde{\mathcal{X}}$ to denote the space of control variables (i.e., SOL1, CQ1, and SQ1), and $\mathcal{X}$ to denote the product space of control and measurement variables (i.e., SOL1, CQ1, SQ1, and Q5).
Additionally we use $T$ to denote the number of total queries that we make in the joint control-measurement domain, and $K$ to denote the number of posterior function samples drawn from our GP model.
We also use $\mathcal{D}_t$ to denote the dataset of queries that have been made by iteration $t$.
In line 8, we introduce an \emph{expected information gain (EIG)} acquisition function, defined as
\begin{equation}
\begin{aligned}
    \text{EIG}_t \mathlarger{(} x \mid \{\tilde{x}_t^k\}_{k=1}^K \mathlarger{)}
    &=\text{H}[ y_x |\mathcal{D}_t] - \mathbb{E}_{p(\tilde{x}^*|\mathcal{D}_t)} [\text{H}[ y_x |\mathcal{D}_t, \tilde{x}^*]]\\
    &\approx\text{H}[ y_x |\mathcal{D}_t]-\frac{1}{K} \sum_{k=1}^K [\text{H}[ y_x |\mathcal{D}_t, \tilde{x}_t^k]],
\end{aligned}
\end{equation}
where $\tilde{x}^*$ denotes the control variable that minimizes the emittance, i.e., $\tilde{x}^* = \text{argmin}_{\tilde{x} \in \tilde{\mathcal{X}}} \hspace{1mm} \text{Emittance}_f(\tilde{x})$, and $\{\tilde{x}_t^k\}_{k=1}^K$ denotes a set of posterior samples of emittance minimizers, generated by running an emittance minimization procedure on posterior function samples, as detailed in Algorithm~\ref{alg:multipointbax} (lines 5-6).
Further, $\text{H}[\hspace{1mm}\cdot\hspace{1mm}]$ denotes the (Shannon) entropy, and $\mathbb{E}_{p(\tilde{x}^*|\mathcal{D}_t)}[\hspace{1mm}\cdot\hspace{1mm}]$ denotes the expectation taken with respect to the posterior density $p(\tilde{x}^*|\mathcal{D}_t)$.
Our acquisition function can be viewed as a version of Bayesian algorithm execution (BAX) \cite{neiswanger2021bayesian}, which aims to greedily gain information about a computed function property via running an algorithm on posterior samples; here, the associated algorithm is an emittance optimization procedure, and the associated function property (computed by the algorithm) is the emittance minimizer over the control variables, $x^* \in \tilde{\mathcal{X}}$.
Our acquisition function is discussed in more detail in \ref{appendixbax}.
Timing results for this algorithm are given in \ref{appendixtiming}.

\begin{algorithm}[H]
    \caption{\textsc{Multipoint-BAX} for Emittance Tuning}
    \label{alg:multipointbax}
    \textbf{Input:} initial dataset $\mathcal{D}_1$, black-box function $f$, prior distribution $p(f)$.
    \begin{algorithmic}[1]
      \For{$t = 1,\ldots,T$}
        \For{$k = 1,\ldots,K$}
            \vspace{0.5mm}
            \State{\textcolor{gray}{\texttt{\# Draw a posterior function sample, then find the control}}}
            \State{\textcolor{gray}{\texttt{\# variable with lowest emittance, as given by this sample}}}
            \State{$\tilde{f}_k \sim p(f \mid \mathcal{D}_t)$}
            \State{$\tilde{x}_t^k \leftarrow \text{argmin}_{\tilde{x} \in \tilde{\mathcal{X}}} \hspace{1mm} \text{Emittance}_{\tilde{f}_k}(\tilde{x})$}
        \EndFor
        \State{\textcolor{gray}{\texttt{\# Optimize the acquisition function}}}
        \State{$x_t \leftarrow \text{argmax}_{x \in \mathcal{X}} \text{EIG}_t(x \mid \{\tilde{x}_t^k\}_{k=1}^K)$}
        \vspace{1mm}
        \State{\textcolor{gray}{\texttt{\# Query black-box function $f$ at $x_t$}}}
        \State $y_{x_t} \sim f(x_t) + \epsilon$
        \vspace{1mm}
        \State{\textcolor{gray}{\texttt{\# Update dataset with new observation}}}
        \State $\mathcal{D}_{t+1} \leftarrow \mathcal{D}_t \cup \{(x_t, y_{x_t})\}$
      \EndFor 
    \end{algorithmic}
    \textbf{Output:} distribution
    $p(f \mid \mathcal{D}_{T+1})$
\end{algorithm}

We note that the multipoint query optimization setting we consider in this paper---and the information-based acquisition function that we develop for this setting---are distinct from the settings (and corresponding methods) of several recent developments in BO, including standard optimization with entropy search methods \cite{hennig2012entropy, hernandez2014predictive, wang2017max}, multi-objective  optimization \cite{belakaria2019max, kandasamy2020tuning}, and multi-fidelity optimization \cite{kandasamy2017multi, belakaria2020multi}.
In particular, multi-fidelity Bayesian optimization (MFBO) settings are related in that they also involve an auxiliary input space (i.e., a \emph{fidelity} space). However, in MFBO, the goal is to optimize the black-box function at the highest fidelity value in the auxiliary space, while in multipoint optimization the goal is to optimize an objective calculated across multiple points in the auxiliary space (for example, via a potentially complex emittance calculation); as a consequence, the goals in the two settings are dissimilar, and methods derived for MFBO do not directly apply to our setting.

\section{Results of Optimization in Noisy LCLS Simulation Setting}
We study the performance of \textsc{Multipoint-BAX} in emittance optimization using a surrogate model of the LCLS copper injector trained on IMPACT-T simulation data. Details of this model are presented in \ref{appendixinjsurr}. At each iteration, we query the beam size from the LCLS surrogate model given a configuration of virtual injector control variables for SOL1, CQ1, and SQ1, as well as a single value for the Q5 strength. For each run, the algorithm was initialized with 10 points that were uniformly sampled from the joint control-measurement domain defined by the bounds set on each variable domain: SOL1: (0.46, 0.485) (kG$\cdot$m), CQ1 and SQ1: (-0.02 0.02) (kG). All tests ran for 200 iterations, corresponding to 200 total beam size queries from the simulation. Each algorithm execution on the posterior (i.e.\ virtual objective) involves a scan of 10 initial points, followed by 10 final points in both $X$ and $Y$ after adjusting the scan range based on the minimum of the parabola (see \ref{appendixemitmeas}). To assess the performance of \textsc{Multipoint-BAX}, we repeat the optimization with black-box methods operating directly on the multipoint query. We chose BO and Nelder-Mead simplex \cite{nelder-mead} as baselines, as both have seen extensive use in accelerator operations \cite{Tomin:2016hnf,McIntire:2016fnl,kirschner2019,hanuka2019,Shalloo_2020,duris2020,Roussel_2021}. Details of the BAX, BO, and simplex algorithm implementations are given in \ref{appendixbax} and \ref{appendixalgo}. 

\begin{figure}
\centering
\hspace{3mm}
\includegraphics[width=0.7\columnwidth]{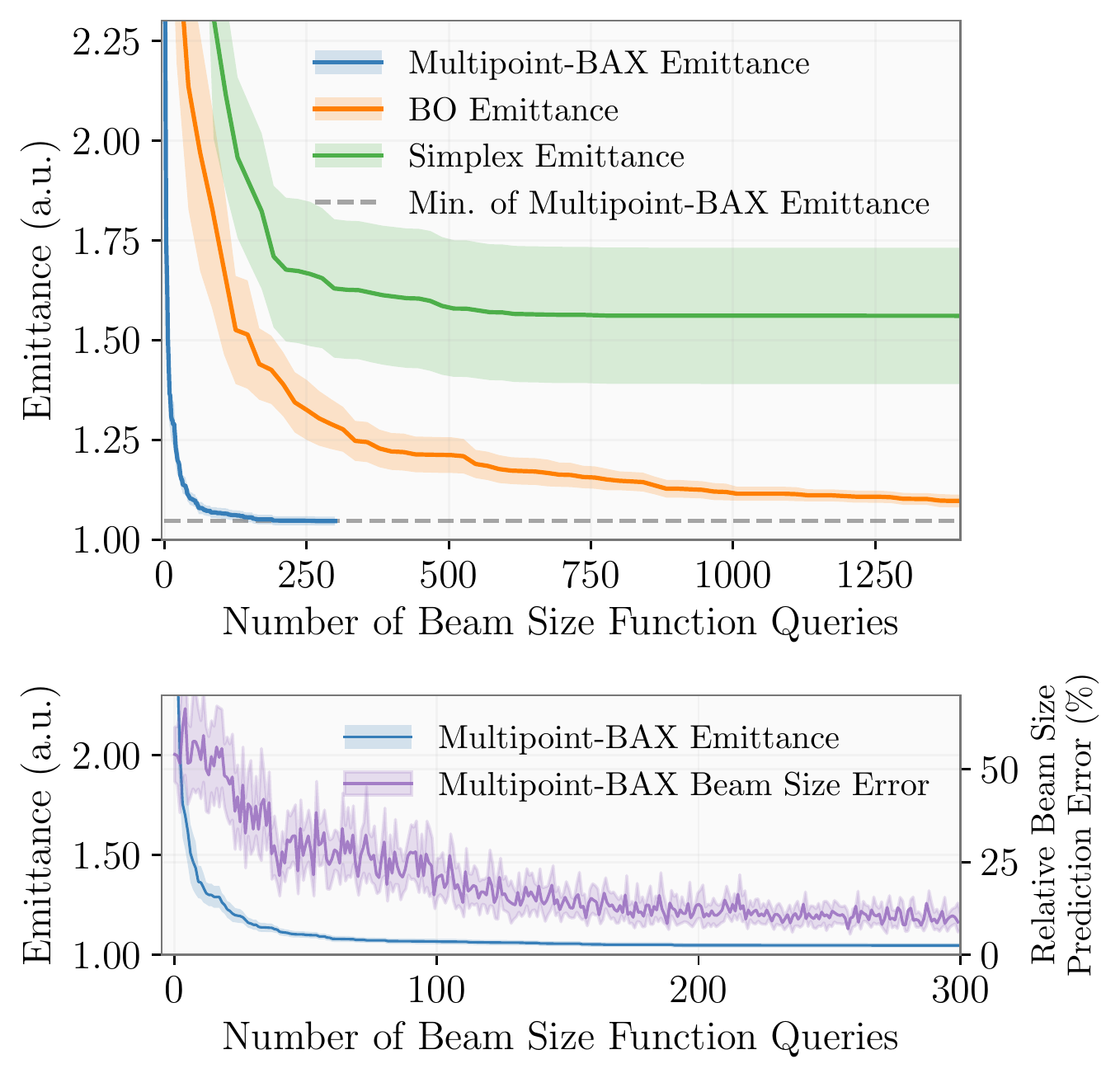}
\caption{\label{fig:sim_results}Simulated LCLS emittance optimization. \textmd{Top: Optimization of beam emittance on the LCLS injector surrogate model using \textsc{Multipoint-BAX} (blue), BO (orange), and Nelder-Mead simplex optimization (green). The mean and 2-sigma standard error of 40 individual runs with randomized starting conditions are shown for each algorithm. The best emittance seen at each iteration is plotted, and the values are normalized by the minimum of the ground truth emittance. Bottom: A magnified view of \textsc{Multipoint-BAX} optimization results (blue), with the error on the beam-size prediction compared to the true (observed) beam size in the LCLS surrogate model are shown on the right-hand axis (purple).}}
\end{figure}

For each algorithm, we ran 40 tests, each with different initial samples. For all simulated measurements, Gaussian noise of the form $\mathcal{N}(0,\sigma_n)$ was added to the beam size for each function query, with $\sigma_n=10$\% as a conservative estimate of typical experimental noise levels at LCLS (see \ref{appendixexpsetup}). Figure \ref{fig:sim_results} shows the mean and standard error of the simulated \textsc{Multipoint-BAX} optimization runs (blue), Bayesian optimization with a GP model (orange), and simplex optimization runs (green). The emittance value shown is the ground truth (noiseless) emittance corresponding to the best control variables found at each iteration, and we display the best emittance up to that point as a function of the total number of beam-size measurements. The results are normalized by the ground truth optimal emittance. In the lower plot of Figure \ref{fig:sim_results}, we show a magnified view of the \textsc{Multipoint-BAX} optimization. A drawback from using a virtual objective is that the algorithm never runs full emittance calculations on the true function during optimization, so in practice an operator would not have access to the blue line. However, we can observe the quality of the internal beam-size model, which can serve as a proxy for convergence; if the GP models the beam-size response correctly, it should also identify the configuration with the optimal emittance. For this reason, we also show the error on its beam-size prediction compared to the true (observed) beam size in the LCLS surrogate model on the right-hand axis (purple). 

\begin{figure}
\centering
\includegraphics[width=0.38\linewidth]{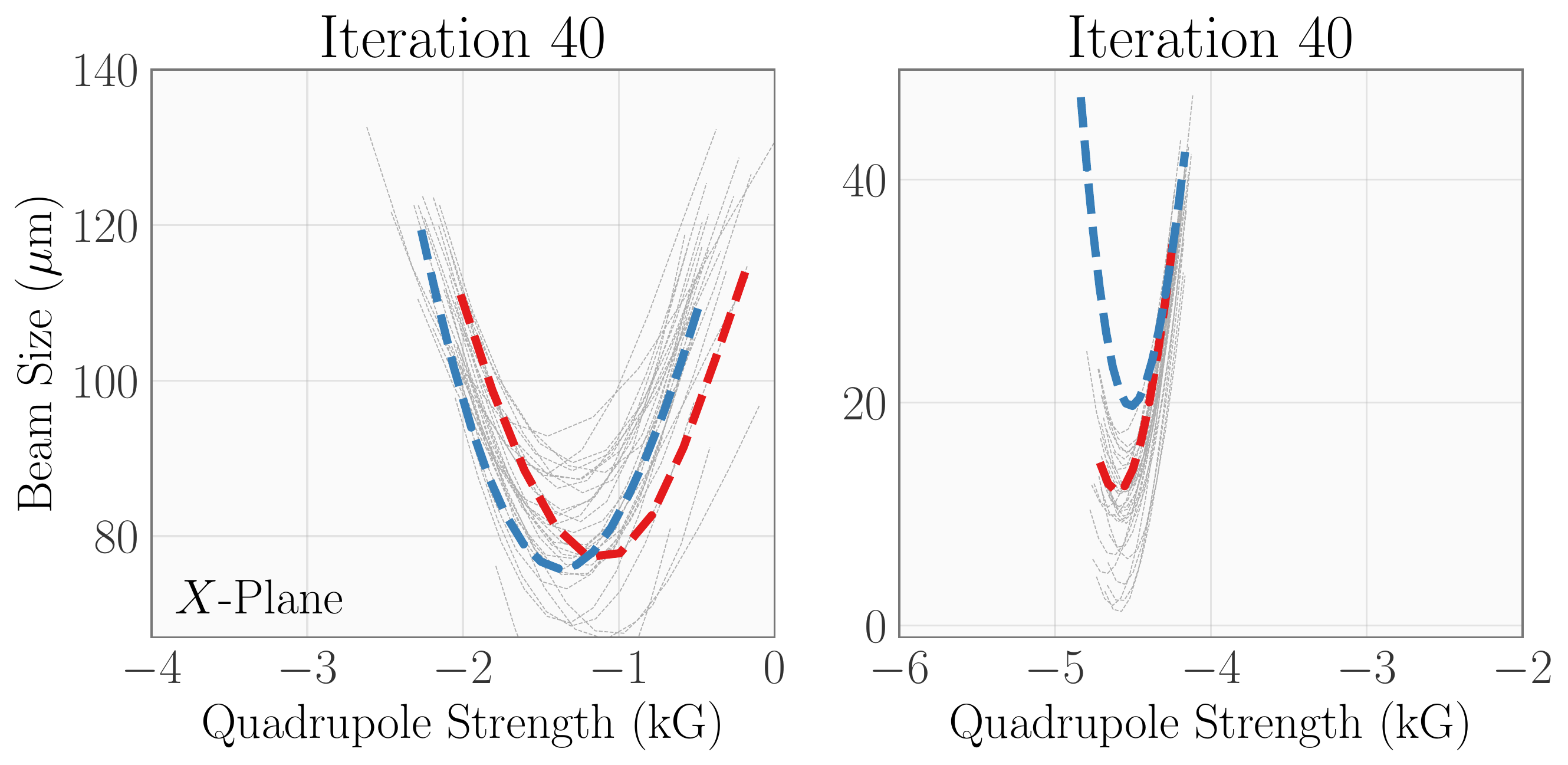}
\hspace{3mm}
\includegraphics[width=0.38\linewidth]{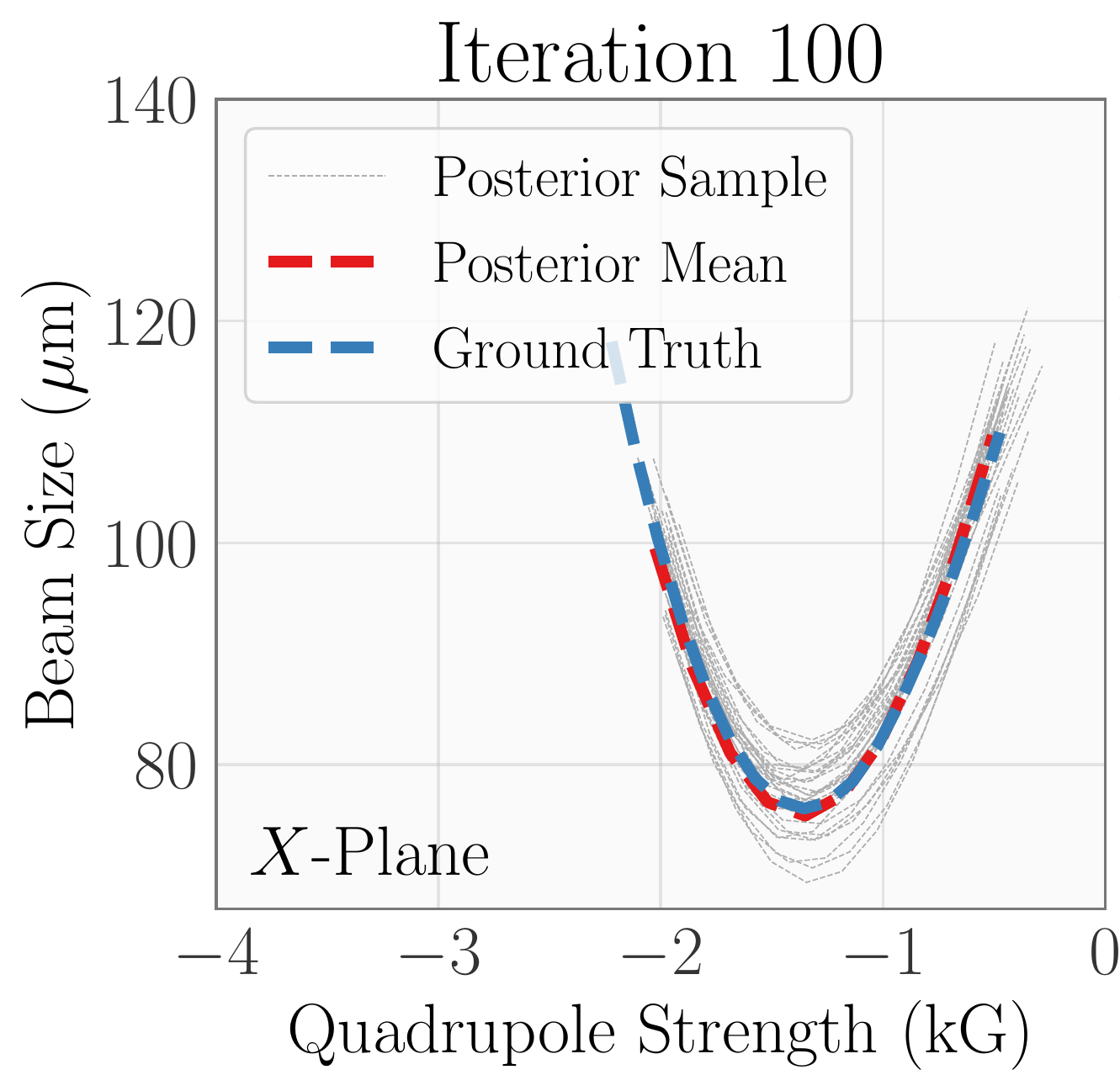}
\caption{\label{fig:sim_posterior} Posterior Samples in \textsc{Multipoint-BAX}. \textmd{Samples of the GP posterior during iterations 40 and 100 during a \textsc{Multipoint-BAX} optimization run on the LCLS surrogate model, with the control variables set to the optimal configuration that the model estimates at the given iteration. We plot posterior samples of the beam-size scan for these control variables (gray), the posterior mean of the GP at the same control variables (red), and the ground truth beam sizes for this scan (blue).}}
\end{figure}

We can further check that the \textsc{Multipoint-BAX} GP model learns the correct behavior of the beam size during scans over the secondary domain. Figure \ref{fig:sim_posterior} shows samples of the GP posterior during iterations 40 and 100, with the control variables set to the optimal configuration that the algorithm estimates at that given iteration. By iteration 100, \textsc{Multipoint-BAX} models the beam-size response to a quadrupole scan with high accuracy.

\textsc{Multipoint-BAX} significantly surpasses BO and simplex in finding an optimal emittance. While each BO and simplex iteration corresponds to an average of 18 beam-size measurements, each algorithm iteration is a single measurement. In this noisy environment, \textsc{Multipoint-BAX} finds the optimal emittance after about 160 measurements on average, while the mean of the BO runs takes over 3000. With each beam-size measurement taking 18 seconds on a wire scanner at LCLS, BAX would require 48 minutes of invasive measurements compared to 15 hours with BO.

\begin{figure}
\centering
\includegraphics[width=0.7\columnwidth]{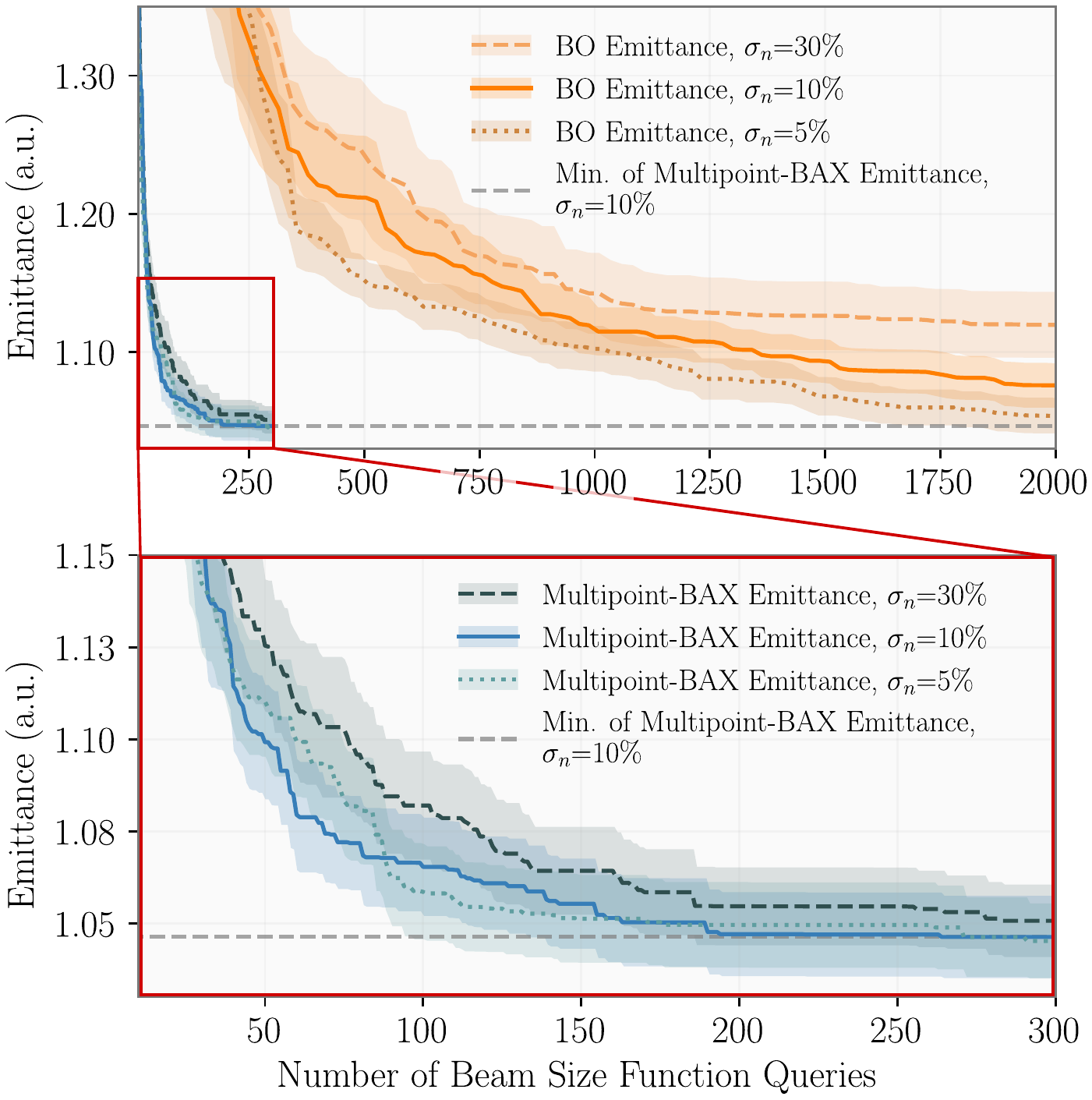}
\caption{\label{fig:sim_noise}Robustness to noise during optimization. \textmd{Comparison of the optimization performance on the LCLS injector surrogate model with varying levels of noise added to the beam size for \textsc{Multipoint-BAX} (blue shades) and BO (orange shades). The best emittance up to that iteration is shown, normalized by the minimum of the ground truth emittance. The lower plot shows a magnified view of \textsc{Multipoint-BAX} optimization.}}
\end{figure}

We hypothesize that \textsc{Multipoint-BAX} should exhibit increased robustness to noise because its GP models the beam size rather than the calculated emittance. Because the emittance fit is sensitive to errors in the beam size, the emittance is more difficult to model than the beam size. To test this robustness, we repeated the previous experiment with both a noisier ($\sigma_n = 30\%$) and a less noisy ($\sigma_n = 5\%$) beam-size function. The results are shown in Figure~\ref{fig:sim_noise}: the increased noise leaves our method's performance largely unchanged (blue), while degrading BO's performance significantly (orange).

\section{Results of Live Experimental Optimization at LCLS and FACET-II}
\label{sec:liveresults}
We applied \textsc{Multipoint-BAX} to the online optimization of the beam emittance in the LCLS injector with an identical setup to the simulated environment. The injector control variables were the solenoid SOL1, and the two quadrupoles CQ1 and SQ1. Emittance calculations scanned the Q5 quadrupole while measuring beam size with a downstream wire scanner. Details of the experimental setup are given in \ref{appendixexpsetup}. The bounds on the SOL1 device domain were identical to the simulation runs, (0.46, 0.485) (kG$\cdot$m), while CQ1 and SQ1's domains were made smaller to (-0.015, 0.015) (kG) due to constraints on the region where beam-size measurements were valid and within range of the wire scanner. In future work, such constraints can be incorporated in the optimization to learn to avoid bad or invalid regions without having to place tight restrictions on the control domains \cite{roussel2021}. 

\begin{figure}
\centering
\hspace{8mm}
\includegraphics[width=0.75\linewidth]{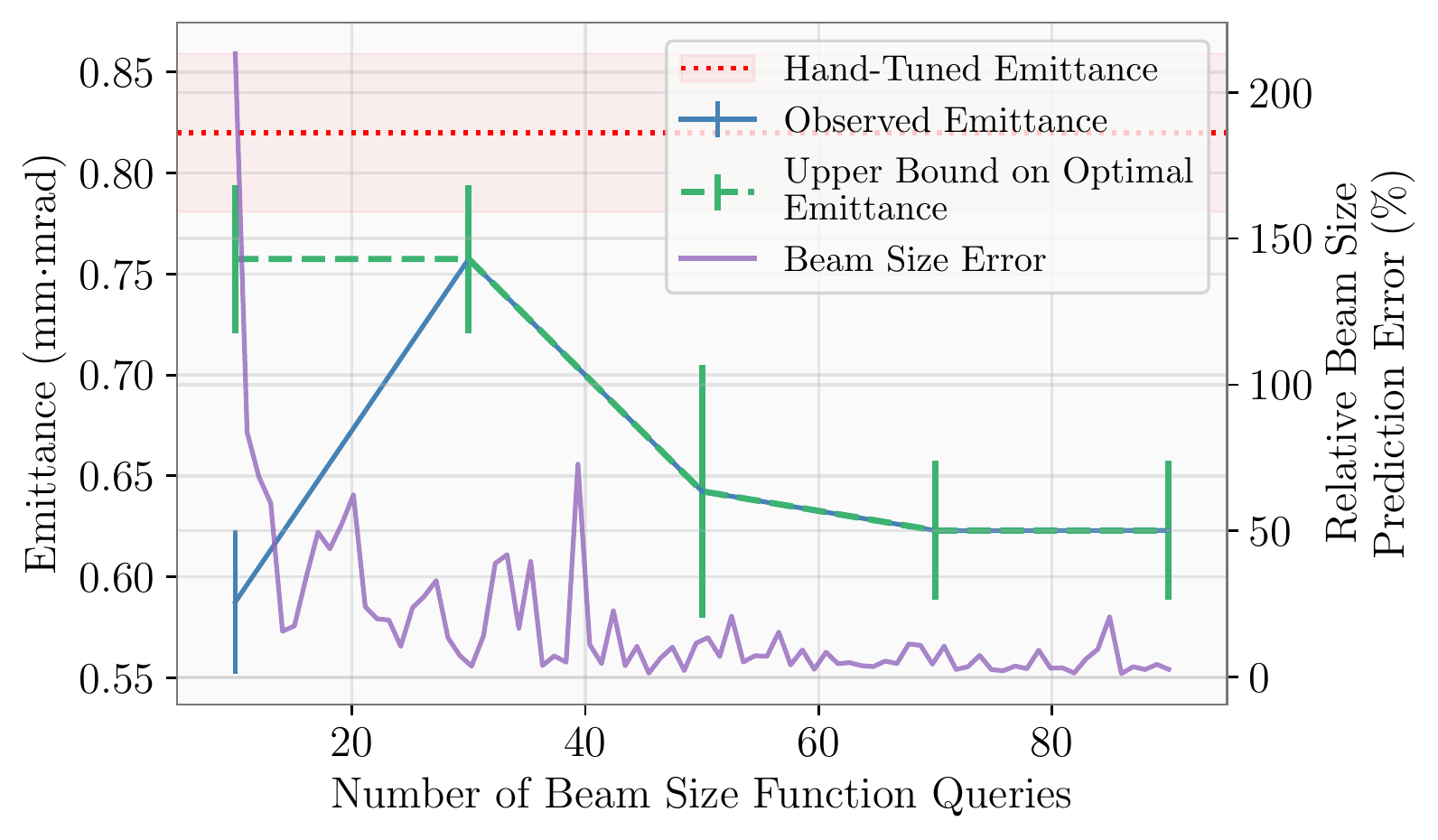}
\caption{\label{fig:lcls_results} Emittance optimization in the LCLS injector. \textmd{Results of the BAX emittance optimization showing the true (observed) emittance calculated from measurements at the optimal configuration identified by \textsc{Multipoint-BAX} every 20 iterations (blue), its upper bound, i.e. the worst emittance seen after that iteration (green), the error on its beam-size prediction compared to the true beam size (purple), and the best hand-tuned emittance on that day for reference (red). Note that the blue and green curves are equal after query 30. See the text in Section~\ref{sec:liveresults} for additional details on each curve shown.}}
\vspace{-2mm}
\end{figure}

Figure \ref{fig:lcls_results} shows the resulting emittance optimization as a function of the number of beam-size measurements, initialized with 10 random points. To assess the performance of \textsc{Multipoint-BAX}, full experimental emittance calculations were performed every 20 iterations at the optimal control variables found by the algorithm at that iteration (blue). The experimental emittance calculations are slow and are only needed to assess convergence for the study. Because an operator would not have access to such calculations in practice, it would not be possible to select an earlier setting with better emittance as would be done with BO. Therefore we show an upper bound on emittance at each point (green), i.e. the worst emittance seen after that iteration. Even without experimental  emittance calculations, an operator can monitor the improvement in the \textsc{Multipoint-BAX} beam-size prediction compared to the measured values (purple) or the settling of control variables to decide when the algorithm has converged. \textsc{Multipoint-BAX} reaches a final emittance of 0.62 $\pm$ 0.02 mm$\cdot$mrad, 24\% lower than that achieved by accelerator operators via hand-tuning during normal operations with the same machine state (0.82 $\pm$ 0.02 mm$\cdot$mrad, in red with the shading corresponding to its 2-sigma error). The matching parameter was 1.17 in the $X$-plane and 1.08 in the $Y$. We note that neither the matching quadrupoles nor the match were included in this optimization.

We can see that \textsc{Multipoint-BAX} converges to small beam-size errors and a low emittance value after approximately 60 iterations, corresponding to about 20 minutes of beam time.
We note that, after only a handful of queries, the model posterior already begins to concentrate on the optimal region, and thus the algorithm has some chance of estimating a low emittance value, as we observe in the blue line in Figure~\ref{fig:lcls_results}.
However, the model error is still large for at least 15 iterations, and yields an inconsistent estimate of the emittance (purple line, Figure~\ref{fig:lcls_results}).
Only after approximately 60 iterations does the model converge to within the measurement error, yielding a final estimate of the emittance minimizer.

\begin{figure}
\centering
\includegraphics[width=0.38\linewidth]{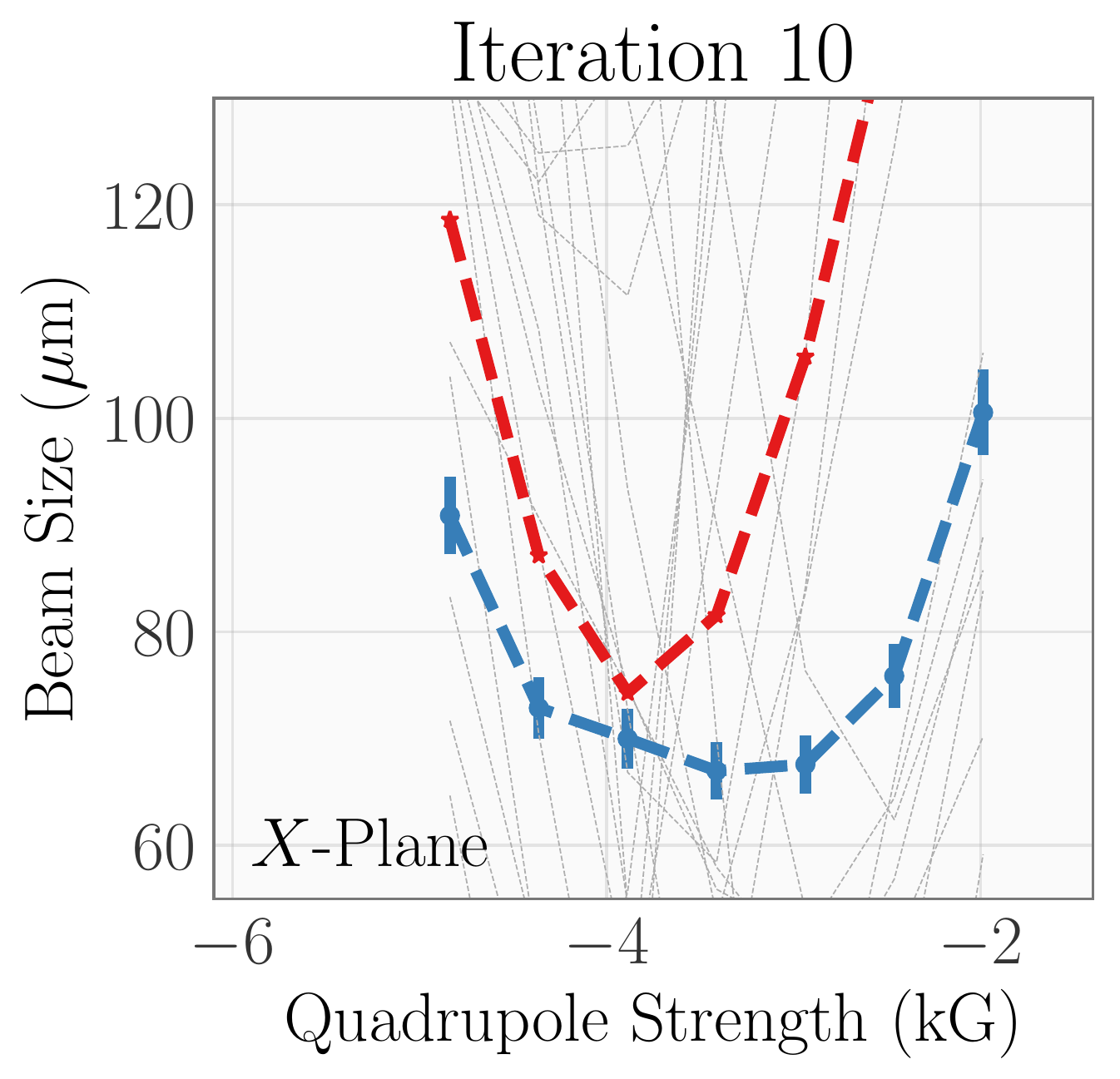}
\hspace{3mm}
\includegraphics[width=0.38\linewidth]{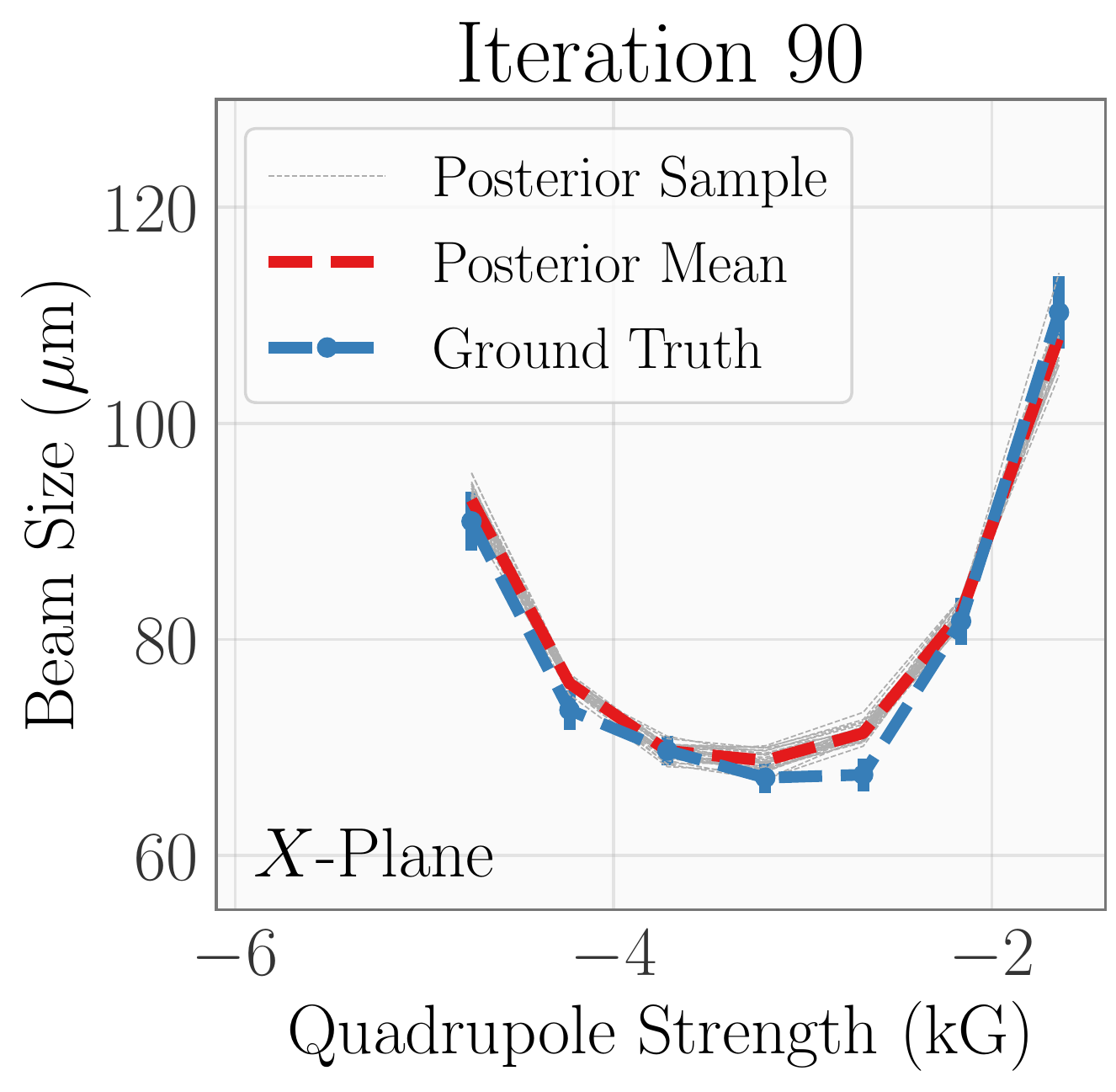}
\caption{\label{fig:lcls_bax_beamsize} \textsc{Multipoint-BAX} beam-size predictions in the LCLS injector. \textmd{Comparison of the beam-size predictions in \textsc{Multipoint-BAX} given by the posterior samples of the beam-size scan (gray) and the posterior mean of the GP (red) at the estimated optimal configuration, and the true observed beam sizes (blue) for a beam-size scan after 10 iterations (left) and 90 iterations (right).}}
\end{figure}

We can also confirm that by the end of the optimization, the GP has learned an accurate model of beam-size scans in the secondary domain, as seen in Figure~\ref{fig:lcls_bax_beamsize}.
For iterations 10 and 90, we plot the posterior samples of the beam-size scan (gray) at the optimal configuration that \textsc{Multipoint-BAX} estimates at that iteration, the posterior mean of the GP (red), and the true beam-size measurements for this scan (blue). We can see that at iteration 10, the algorithm has not yet learned to model beam-size scan accurately, again confirming the low emittance in Figure~\ref{fig:lcls_results} was selected by chance at that iteration. 

The \textsc{Multipoint-BAX} behavior in this experimental setting is consistent with the simulation study on the injector surrogate model. Due to the inefficiency of traditional optimizers, their performance could not be compared to our method on the LCLS machine, as a single BO run was expected to take several hours based on previous experience, and beam time at LCLS is in high demand. Nonetheless, these results are a strong indication that \textsc{Multipoint-BAX} performance surpasses standard black-box optimization methods in noisy experimental settings, just as we observed in the simulation environment.

\begin{figure}
\centering
\hspace{5mm}
\includegraphics[width=0.68\linewidth]{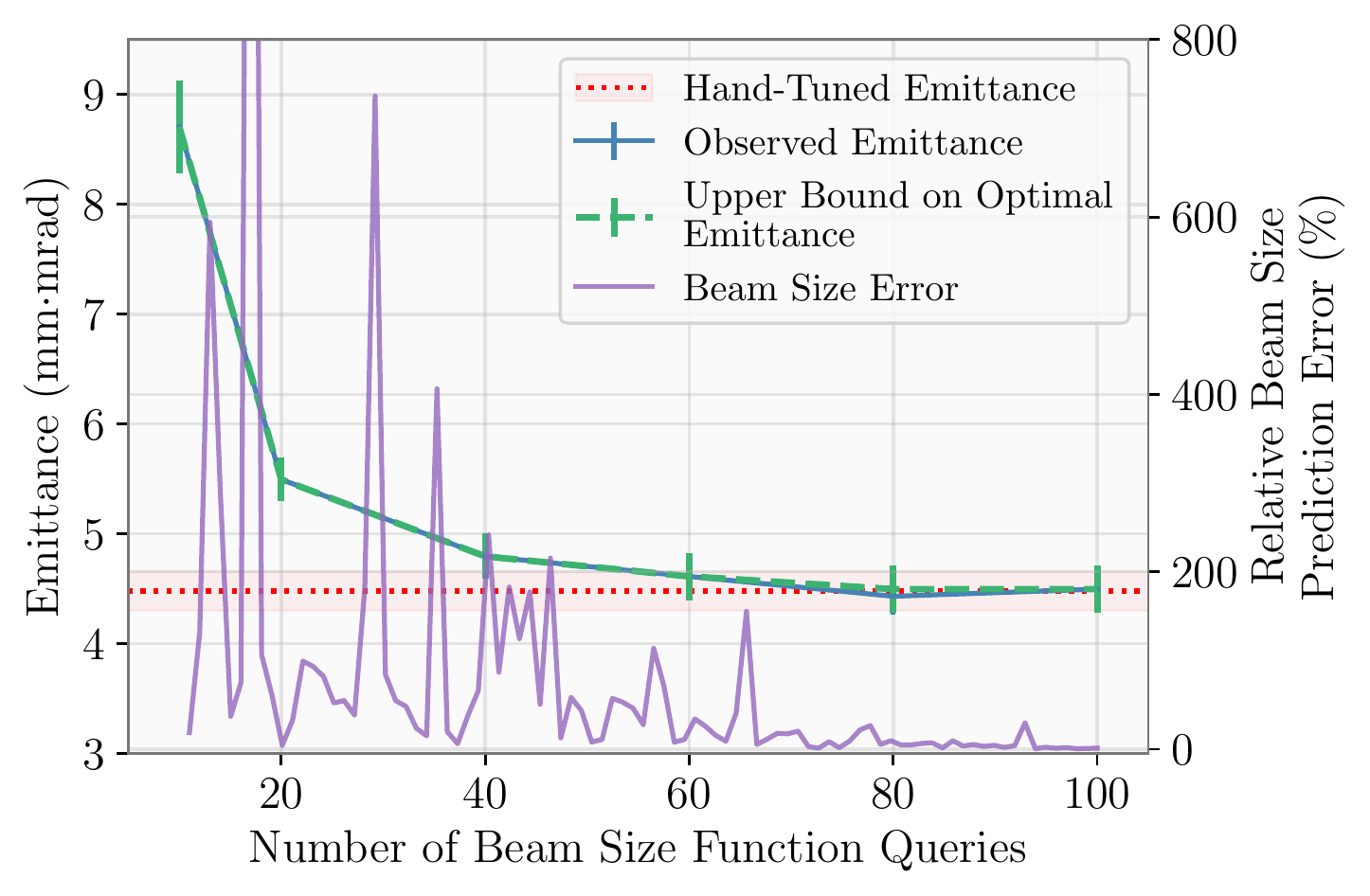}
\caption{\label{fig:facet_results} Emittance optimization in the FACET-II injector. \textmd{Results of the \textsc{Multipoint-BAX} emittance optimization showing the true (observed) emittance calculated from measurements at the optimal configuration identified by the algorithm every 20 iterations (blue), its upper bound, i.e. the worst emittance seen after that iteration (green), the error on the model's beam-size prediction compared to the true beam size (purple), and the best hand-tuned emittance on that day for reference (red). Note that the blue and green curves overlap as they are are equal in this experiment.}}
\end{figure}

In addition to the experimental test on LCLS, we applied the \textsc{Multipoint-BAX} procedure to the optimization of the electron-beam emittance on FACET-II. The FACET-II injector line has an identical setup to the LCLS beam line, and we again optimized SOL1, CQ1 and SQ1. However, FACET-II has a charge of 2 nC, compared to 250 pC in the LCLS run. The control domain bounds were (0.37, 0.41) (kG$\cdot$m) for SOL1, and (-0.01 0.01) (kG) for CQ1 and SQ1. Here, optical transition radiation (OTR) screens were used instead of wire scanners for measuring the beam size downstream of Q5 (see \ref{appendixexpsetup}). 

\textsc{Multipoint-BAX} was initialized with 10 random beam-size measurements, and full emittance evaluations were performed every 20 iterations using the current optimal control variables. Figure \ref{fig:facet_results} shows the results of the FACET-II experimental run.
Following the 10 initial random samples, \textsc{Multipoint-BAX} converges to an optimal configuration after 80 queries, recovering a similar emittance (4.49 $\pm$ 0.11 mm$\cdot$mrad, with a match of 1.07 in $X$-plane and 1.06 in $Y$) to the best hand-tuned value found during normal operations that day (4.48 $\pm$ 0.09 mm$\cdot$mrad, with a match of 1.05 in $X$ and 1.02 in $Y$). Similar to previous figures, we also show the error on the algorithm's beam-size prediction compared to the true beam size measured on the machine at every iteration.

\section{Conclusions and Outlook}
\label{sec:conclusion}
In summary, we have proposed a new approach for highly-efficient optimization of systems involving multipoint queries, and have demonstrated the method experimentally for the high-impact case of emittance optimization in accelerators. Rather than optimizing on slow multipoint queries, \textsc{Multipoint-BAX} instead learns a model of the system on-the-fly from single measurements in the joint control-measurement domain. The algorithm then guides the optimization by calculating a virtual objective on the fast-executing surrogate. The method avoids the need for the full, slow multipoint query while also maximizing the information gain of each measurement, increasing the overall efficiency of the optimization.

We applied \textsc{Multipoint-BAX} to the specific task of electron-beam emittance optimization in the LCLS injector in both noisy simulation and experimental settings, and in the FACET-II injector in an experimental setting. In a simulation with typical LCLS noise levels, we saw a 20$\times$ increase in efficiency in reaching the optimum when using our method compared to BO. In experiments on the live machine, \textsc{Multipoint-BAX} was able to reach an emittance that was 24\% lower than that achieved by hand-tuning at LCLS, and recovered a similar emittance to the best hand-tuned emittance in FACET-II.
As shown in Figure~\ref{fig:lclsii-estimate}, even a modest reduction of emittance at LCLS-II-HE \cite{osti_1634206} can have a dramatic impact on FEL performance.

Future work will include expanding the \textsc{Multipoint-BAX} method presented here to higher dimensions to incorporate more quadrupoles and control variables along the accelerator, and targeting more complex objectives (e.g. the beam matching parameter along with the emittance). We expect performance can improve further with stronger priors, such as information on the physical correlations between the control and measurement variables \cite{duris2020}. 
Using a non-zero prior mean function, e.g. a surrogate learned from simulations or previous measurements, may also improve tuning efficiency \cite{xu2022}. This method could also be used in tandem with more comprehensive machine models, and highlights a new path toward replacing expensive indirect beam measurements with computation on easy-to-acquire samples from surrogate models.
In addition to emittance minimization, a variety of other accelerator tasks require multipoint queries and could employ \textsc{Multipoint-BAX}, for example manipulating longitudinal phase space \cite{PhysRevAccelBeams.23.114201} or designing X-ray focusing optics.
Beyond accelerators, we anticipate \textsc{Multipoint-BAX} will be broadly applicable to complex optimization problems in science and engineering. While we focus on multipoint queries, similar advantages exist for single-point queries that produce rich outputs, e.g. from simulations. \textsc{Multipoint-BAX} is particularly advantageous over traditional methods on tasks with high noise environments and high dimensional inputs, both common to real-world scientific instruments.

\section*{Acknowledgements}
This work is supported by the U.S. Department of Energy, Office of Science, Office of Basic Energy Sciences under Contract No. DE-AC02-76SF00515. The authors would like to thank Nicole Neveu for her help in setting up the LCLS emittance calculation, Finn O'Shea for his feedback on the emittance calculation procedures, and Feng Zhou and John Sheppard for their guidance and support with the LCLS injector machine development time.

\newpage
\bibliographystyle{unsrtnat}
\bibliography{bax}

\begin{thebibliography}{64}
\providecommand{\natexlab}[1]{#1}
\providecommand{\url}[1]{\texttt{#1}}
\expandafter\ifx\csname urlstyle\endcsname\relax
  \providecommand{\doi}[1]{doi: #1}\else
  \providecommand{\doi}{doi: \begingroup \urlstyle{rm}\Url}\fi

\bibitem[Jones et~al.(1998)Jones, Schonlau, and Welch]{jones1998efficient}
Donald~R Jones, Matthias Schonlau, and William~J Welch.
\newblock Efficient global optimization of expensive black-box functions.
\newblock \emph{Journal of Global optimization}, 13\penalty0 (4):\penalty0
  455--492, 1998.

\bibitem[Brochu et~al.(2010)Brochu, Cora, and de~Freitas]{brochu2010}
Eric Brochu, Vlad~M. Cora, and Nando de~Freitas.
\newblock A tutorial on bayesian optimization of expensive cost functions, with
  application to active user modeling and hierarchical reinforcement learning.
\newblock \emph{CoRR}, abs/1012.2599, 2010.
\newblock URL \url{http://arxiv.org/abs/1012.2599}.

\bibitem[Huang and Safranek(2015)]{rcds_demo}
Xiaobiao Huang and James Safranek.
\newblock Online optimization of storage ring nonlinear beam dynamics.
\newblock \emph{Phys. Rev. ST Accel. Beams}, 18:\penalty0 084001, Aug 2015.
\newblock \doi{10.1103/PhysRevSTAB.18.084001}.
\newblock URL \url{https://link.aps.org/doi/10.1103/PhysRevSTAB.18.084001}.

\bibitem[Huang(2018{\natexlab{a}})]{rcds}
Xiaobiao Huang.
\newblock Robust simplex algorithm for online optimization.
\newblock \emph{Phys. Rev. Accel. Beams}, 21:\penalty0 104601, Oct
  2018{\natexlab{a}}.
\newblock \doi{10.1103/PhysRevAccelBeams.21.104601}.
\newblock URL
  \url{https://link.aps.org/doi/10.1103/PhysRevAccelBeams.21.104601}.

\bibitem[Scheinker et~al.(2013)Scheinker, Pang, and Rybarcyk]{ES}
Alexander Scheinker, Xiaoying Pang, and Larry Rybarcyk.
\newblock Model-independent particle accelerator tuning.
\newblock \emph{Phys. Rev. ST Accel. Beams}, 16:\penalty0 102803, Oct 2013.
\newblock \doi{10.1103/PhysRevSTAB.16.102803}.
\newblock URL \url{https://link.aps.org/doi/10.1103/PhysRevSTAB.16.102803}.

\bibitem[Huang(2018{\natexlab{b}})]{HuangRSimplex}
Xiaobiao Huang.
\newblock Robust simplex algorithm for online optimization.
\newblock \emph{Phys. Rev. Accel. Beams}, 21:\penalty0 104601, Oct
  2018{\natexlab{b}}.
\newblock \doi{10.1103/PhysRevAccelBeams.21.104601}.
\newblock URL
  \url{https://link.aps.org/doi/10.1103/PhysRevAccelBeams.21.104601}.

\bibitem[{Scheinker} et~al.(2018){Scheinker}, {Huang}, and
  {Wu}]{ScheinkerSP32018}
A.~{Scheinker}, X.~{Huang}, and J.~{Wu}.
\newblock Minimization of betatron oscillations of electron beam injected into
  a time-varying lattice via extremum seeking.
\newblock \emph{IEEE Transactions on Control Systems Technology}, 26\penalty0
  (1):\penalty0 336--343, Jan 2018.
\newblock ISSN 1063-6536.
\newblock \doi{10.1109/TCST.2017.2664728}.

\bibitem[Bergan et~al.(2019)Bergan, Bazarov, Duncan, Liarte, Rubin, and
  Sethna]{BerganGA2019}
W.~F. Bergan, I.~V. Bazarov, C.~J.~R. Duncan, D.~B. Liarte, D.~L. Rubin, and
  J.~P. Sethna.
\newblock Online storage ring optimization using dimension-reduction and
  genetic algorithms.
\newblock \emph{Phys. Rev. Accel. Beams}, 22:\penalty0 054601, May 2019.
\newblock \doi{10.1103/PhysRevAccelBeams.22.054601}.
\newblock URL
  \url{https://link.aps.org/doi/10.1103/PhysRevAccelBeams.22.054601}.

\bibitem[Scheinker et~al.(2019)Scheinker, Bohler, Tomin, Kammering, Zagorodnov,
  Schlarb, Scholz, Beutner, and Decking]{ES2019}
Alexander Scheinker, Dorian Bohler, Sergey Tomin, Raimund Kammering, Igor
  Zagorodnov, Holger Schlarb, Matthias Scholz, Bolko Beutner, and Winfried
  Decking.
\newblock Model-independent tuning for maximizing free electron laser pulse
  energy.
\newblock \emph{Phys. Rev. Accel. Beams}, 22:\penalty0 082802, Aug 2019.
\newblock \doi{10.1103/PhysRevAccelBeams.22.082802}.
\newblock URL
  \url{https://link.aps.org/doi/10.1103/PhysRevAccelBeams.22.082802}.

\bibitem[Terayama et~al.(2021)Terayama, Sumita, Tamura, and
  Tsuda]{terayama2021black}
Kei Terayama, Masato Sumita, Ryo Tamura, and Koji Tsuda.
\newblock Black-box optimization for automated discovery.
\newblock \emph{Accounts of Chemical Research}, 54\penalty0 (6):\penalty0
  1334--1346, 2021.

\bibitem[Char et~al.(2019)Char, Chung, Neiswanger, Kandasamy, Nelson, Boyer,
  Kolemen, and Schneider]{char2019offline}
Ian Char, Youngseog Chung, Willie Neiswanger, Kirthevasan Kandasamy, Andrew~O
  Nelson, Mark Boyer, Egemen Kolemen, and Jeff Schneider.
\newblock Offline contextual bayesian optimization.
\newblock \emph{Advances in Neural Information Processing Systems}, 32, 2019.

\bibitem[Ueno et~al.(2016)Ueno, Rhone, Hou, Mizoguchi, and
  Tsuda]{ueno2016combo}
Tsuyoshi Ueno, Trevor~David Rhone, Zhufeng Hou, Teruyasu Mizoguchi, and Koji
  Tsuda.
\newblock Combo: An efficient bayesian optimization library for materials
  science.
\newblock \emph{Materials discovery}, 4:\penalty0 18--21, 2016.

\bibitem[Kushner(1963)]{kushner1963new}
Harold~J Kushner.
\newblock A new method of locating the maximum point of an arbitrary multipeak
  curve in the presence of noise.
\newblock In \emph{Joint Automatic Control Conference}, number~1, pages 69--79,
  1963.

\bibitem[Mo{\v{c}}kus(1975)]{mockus1975}
J.~Mo{\v{c}}kus.
\newblock On bayesian methods for seeking the extremum.
\newblock In G.~I. Marchuk, editor, \emph{Optimization Techniques IFIP
  Technical Conference Novosibirsk, July 1--7, 1974}, pages 400--404, Berlin,
  Heidelberg, 1975. Springer Berlin Heidelberg.
\newblock ISBN 978-3-540-37497-8.

\bibitem[McIntire et~al.(2016)McIntire, Cope, Ermon, and
  Ratner]{McIntire:2016fnl}
Mitchell McIntire, Tyler Cope, Stefano Ermon, and Daniel Ratner.
\newblock {Bayesian Optimization of FEL Performance at LCLS}.
\newblock In \emph{{7th International Particle Accelerator Conference}}, page
  WEPOW055, 2016.
\newblock \doi{10.18429/JACoW-IPAC2016-WEPOW055}.

\bibitem[Kirschner et~al.(2019)Kirschner, Mutný, Hiller, Ischebeck, and
  Krause]{kirschner2019}
Johannes Kirschner, Mojmír Mutný, Nicole Hiller, Rasmus Ischebeck, and
  Andreas Krause.
\newblock Adaptive and safe bayesian optimization in high dimensions via
  one-dimensional subspaces, 2019.
\newblock URL \url{https://arxiv.org/abs/1902.03229}.

\bibitem[Duris et~al.(2020)Duris, Kennedy, Hanuka, Shtalenkova, Edelen,
  Baxevanis, Egger, Cope, McIntire, Ermon, and Ratner]{duris2020}
J.~Duris, D.~Kennedy, A.~Hanuka, J.~Shtalenkova, A.~Edelen, P.~Baxevanis,
  A.~Egger, T.~Cope, M.~McIntire, S.~Ermon, and D.~Ratner.
\newblock Bayesian optimization of a free-electron laser.
\newblock \emph{Phys. Rev. Lett.}, 124:\penalty0 124801, Mar 2020.
\newblock \doi{10.1103/PhysRevLett.124.124801}.
\newblock URL \url{https://link.aps.org/doi/10.1103/PhysRevLett.124.124801}.

\bibitem[Shalloo et~al.(2020)Shalloo, Dann, Gruse, Underwood, Antoine, Arran,
  Backhouse, Baird, Balcazar, Bourgeois, Cardarelli, Hatfield, Kang,
  Krushelnick, Mangles, Murphy, Lu, Osterhoff, P{\~{o}}der, Rajeev, Ridgers,
  Rozario, Selwood, Shahani, Symes, Thomas, Thornton, Najmudin, and
  Streeter]{Shalloo_2020}
R.~J. Shalloo, S.~J.~D. Dann, J.-N. Gruse, C.~I.~D. Underwood, A.~F. Antoine,
  C.~Arran, M.~Backhouse, C.~D. Baird, M.~D. Balcazar, N.~Bourgeois, J.~A.
  Cardarelli, P.~Hatfield, J.~Kang, K.~Krushelnick, S.~P.~D. Mangles, C.~D.
  Murphy, N.~Lu, J.~Osterhoff, K.~P{\~{o}}der, P.~P. Rajeev, C.~P. Ridgers,
  S.~Rozario, M.~P. Selwood, A.~J. Shahani, D.~R. Symes, A.~G.~R. Thomas,
  C.~Thornton, Z.~Najmudin, and M.~J.~V. Streeter.
\newblock Automation and control of laser wakefield accelerators using bayesian
  optimization.
\newblock \emph{Nature Communications}, 11\penalty0 (1), dec 2020.
\newblock \doi{10.1038/s41467-020-20245-6}.
\newblock URL \url{https://doi.org/10.1038%2Fs41467-020-20245-6}.

\bibitem[Roussel et~al.(2021{\natexlab{a}})Roussel, Hanuka, and
  Edelen]{Roussel_2021}
Ryan Roussel, Adi Hanuka, and Auralee Edelen.
\newblock Multiobjective bayesian optimization for online accelerator tuning.
\newblock \emph{Phys. Rev. Accel. Beams}, 24:\penalty0 062801, Jun
  2021{\natexlab{a}}.
\newblock \doi{10.1103/PhysRevAccelBeams.24.062801}.
\newblock URL
  \url{https://link.aps.org/doi/10.1103/PhysRevAccelBeams.24.062801}.

\bibitem[Yamashita et~al.(2018)Yamashita, Sato, Kino, Miyake, Tsuda, and
  Oguchi]{yamashita}
Tomoki Yamashita, Nobuya Sato, Hiori Kino, Takashi Miyake, Koji Tsuda, and
  Tamio Oguchi.
\newblock Crystal structure prediction accelerated by bayesian optimization.
\newblock \emph{Phys. Rev. Materials}, 2:\penalty0 013803, Jan 2018.
\newblock \doi{10.1103/PhysRevMaterials.2.013803}.
\newblock URL \url{https://link.aps.org/doi/10.1103/PhysRevMaterials.2.013803}.

\bibitem[Miskovich et~al.(2022{\natexlab{a}})Miskovich, Montes, Berg, Blackmon,
  Chipps, Couder, Deibel, Hermansen, Hood, Jain, Ruland, Schatz, Smith,
  Tsintari, and Wagner]{miskovich}
S.~A. Miskovich, F.~Montes, G.~P.~A. Berg, J.~Blackmon, K.~A. Chipps,
  M.~Couder, C.~M. Deibel, K.~Hermansen, A.~A. Hood, R.~Jain, T.~Ruland,
  H.~Schatz, M.~S. Smith, P.~Tsintari, and L.~Wagner.
\newblock Online bayesian optimization for a recoil mass separator.
\newblock \emph{Phys. Rev. Accel. Beams}, 25:\penalty0 044601, Apr
  2022{\natexlab{a}}.
\newblock \doi{10.1103/PhysRevAccelBeams.25.044601}.
\newblock URL
  \url{https://link.aps.org/doi/10.1103/PhysRevAccelBeams.25.044601}.

\bibitem[Liem et~al.(2015)Liem, Kenway, and Martins]{liem2015}
Rhea~P. Liem, Gaetan K.~W. Kenway, and Joaquim R. R.~A. Martins.
\newblock Multimission aircraft fuel-burn minimization via multipoint
  aerostructural optimization.
\newblock \emph{AIAA Journal}, 53\penalty0 (1):\penalty0 104--122, 2015.
\newblock \doi{10.2514/1.J052940}.
\newblock URL \url{https://doi.org/10.2514/1.J052940}.

\bibitem[Terayama et~al.(2017)Terayama, Iwata, Araki, Okuno, and
  Tsuda]{terayama}
Kei Terayama, Hiroaki Iwata, Mitsugu Araki, Yasushi Okuno, and Koji Tsuda.
\newblock {Machine learning accelerates MD-based binding pose prediction
  between ligands and proteins}.
\newblock \emph{Bioinformatics}, 34\penalty0 (5):\penalty0 770--778, 10 2017.
\newblock ISSN 1367-4803.
\newblock \doi{10.1093/bioinformatics/btx638}.
\newblock URL \url{https://doi.org/10.1093/bioinformatics/btx638}.

\bibitem[Lauber et~al.(2020)Lauber, Aulenbacher, Barth, Dziuba, List, Burandt,
  Gettmann, K\"urzeder, Miski-Oglu, Forck, Heilmann, Rubin, Sieber, Yaramyshev,
  Podlech, and Schwarz]{PhysRevAccelBeams.23.114201}
S.~Lauber, K.~Aulenbacher, W.~Barth, F.~Dziuba, J.~List, C.~Burandt,
  V.~Gettmann, T.~K\"urzeder, M.~Miski-Oglu, P.~Forck, M.~Heilmann, A.~Rubin,
  T.~Sieber, S.~Yaramyshev, H.~Podlech, and M.~Schwarz.
\newblock Longitudinal phase space reconstruction for a heavy ion accelerator.
\newblock \emph{Phys. Rev. Accel. Beams}, 23:\penalty0 114201, Nov 2020.
\newblock \doi{10.1103/PhysRevAccelBeams.23.114201}.
\newblock URL
  \url{https://link.aps.org/doi/10.1103/PhysRevAccelBeams.23.114201}.

\bibitem[Minty and Zimmermann(2003)]{minty}
M.~Minty and F.~Zimmermann.
\newblock \emph{Measurement and Control of Charged Particle Beams}.
\newblock Springer Berlin Heidelberg, 01 2003.
\newblock ISBN 978-3-642-07914-6.
\newblock \doi{10.1007/978-3-662-08581-3}.

\bibitem[Emma et~al.(2010)Emma, Akre, Arthur, Bionta, Bostedt, Bozek,
  Brachmann, Bucksbaum, Coffee, Decker, Ding, Dowell, Edstrom, Fisher, Frisch,
  Gilevich, Hastings, Hays, Hering, Huang, Iverson, Loos, Messerschmidt,
  Miahnahri, Moeller, Nuhn, Pile, Ratner, Rzepiela, Schultz, Smith, Stefan,
  Tompkins, Turner, Welch, White, Wu, Yocky, and Galayda]{Emma2010}
P.~Emma, R.~Akre, J.~Arthur, R.~Bionta, C.~Bostedt, J.~Bozek, A.~Brachmann,
  P.~Bucksbaum, R.~Coffee, F.-J. Decker, Y.~Ding, D.~Dowell, S.~Edstrom,
  A.~Fisher, J.~Frisch, S.~Gilevich, J.~Hastings, G.~Hays, Ph. Hering,
  Z.~Huang, R.~Iverson, H.~Loos, M.~Messerschmidt, A.~Miahnahri, S.~Moeller,
  H.-D. Nuhn, G.~Pile, D.~Ratner, J.~Rzepiela, D.~Schultz, T.~Smith, P.~Stefan,
  H.~Tompkins, J.~Turner, J.~Welch, W.~White, J.~Wu, G.~Yocky, and J.~Galayda.
\newblock First lasing and operation of an {\aa}ngstrom-wavelength
  free-electron laser.
\newblock \emph{Nature Photonics}, 4\penalty0 (9):\penalty0 641--647, August
  2010.
\newblock \doi{10.1038/nphoton.2010.176}.
\newblock URL \url{https://doi.org/10.1038/nphoton.2010.176}.

\bibitem[Huang and Kim(2007)]{kim2007}
Zhirong Huang and Kwang-Je Kim.
\newblock Review of x-ray free-electron laser theory.
\newblock \emph{Phys. Rev. ST Accel. Beams}, 10:\penalty0 034801, Mar 2007.
\newblock \doi{10.1103/PhysRevSTAB.10.034801}.
\newblock URL \url{https://link.aps.org/doi/10.1103/PhysRevSTAB.10.034801}.

\bibitem[Schoenlein(2016)]{osti_1634206}
Robert~W Schoenlein.
\newblock Lcls-ii high energy (lcls-ii-he): A transformative x-ray laser for
  science.
\newblock 1 2016.
\newblock \doi{10.2172/1634206}.
\newblock URL \url{https://www.osti.gov/biblio/1634206}.

\bibitem[Brinkmann et~al.(2001)Brinkmann, Derbenev, and
  Fl\"ottmann]{PhysRevSTAB.4.053501}
R.~Brinkmann, Y.~Derbenev, and K.~Fl\"ottmann.
\newblock A low emittance, flat-beam electron source for linear colliders.
\newblock \emph{Phys. Rev. ST Accel. Beams}, 4:\penalty0 053501, May 2001.
\newblock \doi{10.1103/PhysRevSTAB.4.053501}.
\newblock URL \url{https://link.aps.org/doi/10.1103/PhysRevSTAB.4.053501}.

\bibitem[Benedikt et~al.(2015)Benedikt, Schulte, and
  Zimmermann]{PhysRevSTAB.18.101002}
Michael Benedikt, Daniel Schulte, and Frank Zimmermann.
\newblock Optimizing integrated luminosity of future hadron colliders.
\newblock \emph{Phys. Rev. ST Accel. Beams}, 18:\penalty0 101002, Oct 2015.
\newblock \doi{10.1103/PhysRevSTAB.18.101002}.
\newblock URL \url{https://link.aps.org/doi/10.1103/PhysRevSTAB.18.101002}.

\bibitem[Piot et~al.(2006)Piot, Sun, and Kim]{PhysRevSTAB.9.031001}
P.~Piot, Y.-E Sun, and K.-J. Kim.
\newblock Photoinjector generation of a flat electron beam with transverse
  emittance ratio of 100.
\newblock \emph{Phys. Rev. ST Accel. Beams}, 9:\penalty0 031001, Mar 2006.
\newblock \doi{10.1103/PhysRevSTAB.9.031001}.
\newblock URL \url{https://link.aps.org/doi/10.1103/PhysRevSTAB.9.031001}.

\bibitem[Ody et~al.(2017)Ody, Musumeci, Maxson, Cesar, England, and
  Wootton]{ODY201775}
A.~Ody, P.~Musumeci, J.~Maxson, D.~Cesar, R.J. England, and K.P. Wootton.
\newblock Flat electron beam sources for dla accelerators.
\newblock \emph{Nuclear Instruments and Methods in Physics Research Section A:
  Accelerators, Spectrometers, Detectors and Associated Equipment},
  865:\penalty0 75--83, 2017.
\newblock ISSN 0168-9002.
\newblock \doi{https://doi.org/10.1016/j.nima.2016.10.041}.
\newblock URL
  \url{https://www.sciencedirect.com/science/article/pii/S0168900216310877}.
\newblock Physics and Applications of High Brightness Beams 2016.

\bibitem[Neiswanger et~al.(2021)Neiswanger, Wang, and
  Ermon]{neiswanger2021bayesian}
Willie Neiswanger, Ke~Alexander Wang, and Stefano Ermon.
\newblock Bayesian algorithm execution: Estimating computable properties of
  black-box functions using mutual information.
\newblock In \emph{International Conference on Machine Learning}. PMLR, 2021.

\bibitem[Nelder and Mead(1965)]{nelder-mead}
J.~A. Nelder and R.~Mead.
\newblock {A Simplex Method for Function Minimization}.
\newblock \emph{The Computer Journal}, 7\penalty0 (4):\penalty0 308--313, 01
  1965.
\newblock ISSN 0010-4620.
\newblock \doi{10.1093/comjnl/7.4.308}.
\newblock URL \url{https://doi.org/10.1093/comjnl/7.4.308}.

\bibitem[Yakimenko et~al.(2019)Yakimenko, Alsberg, Bong, Bouchard, Clarke,
  Emma, Green, Hast, Hogan, Seabury, Lipkowitz, O'Shea, Storey, White, and
  Yocky]{yakimenko2018}
V.~Yakimenko, L.~Alsberg, E.~Bong, G.~Bouchard, C.~Clarke, C.~Emma, S.~Green,
  C.~Hast, M.~J. Hogan, J.~Seabury, N.~Lipkowitz, B.~O'Shea, D.~Storey,
  G.~White, and G.~Yocky.
\newblock Facet-ii facility for advanced accelerator experimental tests.
\newblock \emph{Phys. Rev. Accel. Beams}, 22:\penalty0 101301, Oct 2019.
\newblock \doi{10.1103/PhysRevAccelBeams.22.101301}.
\newblock URL
  \url{https://link.aps.org/doi/10.1103/PhysRevAccelBeams.22.101301}.

\bibitem[Wang et~al.(1991)Wang, Wang, and Reiser]{WANG1991190}
J.G. Wang, D.X. Wang, and M.~Reiser.
\newblock Beam emittance measurement by the pepper-pot method.
\newblock \emph{Nuclear Instruments and Methods in Physics Research Section A:
  Accelerators, Spectrometers, Detectors and Associated Equipment},
  307\penalty0 (2):\penalty0 190--194, 1991.
\newblock ISSN 0168-9002.
\newblock \doi{https://doi.org/10.1016/0168-9002(91)90182-P}.
\newblock URL
  \url{https://www.sciencedirect.com/science/article/pii/016890029190182P}.

\bibitem[{Thangaraj} and {Piot}(2012)]{2012AIPC.1507..757T}
J.~C.~T. {Thangaraj} and P.~{Piot}.
\newblock {A high-resolution multi-slit phase space measurement technique for
  low-emittance beams}.
\newblock In Rafal {Zgadzaj}, Erhard {Gaul}, and Michael~C. {Downer}, editors,
  \emph{Advanced Accelerator Concepts: 15th Advanced Accelerator Concepts
  Workshop}, volume 1507 of \emph{American Institute of Physics Conference
  Series}, pages 757--761, December 2012.
\newblock \doi{10.1063/1.4773793}.

\bibitem[Zhang(1996)]{Zhang:1996af}
M.~Zhang.
\newblock Emittance formula for slits and pepper-pot measurement.
\newblock 1996.
\newblock \doi{10.2172/395453}.
\newblock URL \url{https://www.osti.gov/biblio/395453}.

\bibitem[Strehl(2006)]{Strehl2006}
Peter Strehl.
\newblock \emph{Beam Instrumentation and Diagnostics: Measurements in Phase
  Spaces}, pages 213--283.
\newblock Springer Berlin Heidelberg, Berlin, Heidelberg, 2006.
\newblock ISBN 978-3-540-26404-0.
\newblock \doi{10.1007/3-540-26404-3_6}.
\newblock URL \url{https://doi.org/10.1007/3-540-26404-3_6}.

\bibitem[Akre et~al.(2008)Akre, Dowell, Emma, Frisch, Gilevich, Hays, Hering,
  Iverson, Limborg-Deprey, Loos, Miahnahri, Schmerge, Turner, Welch, White, and
  Wu]{lclsinj}
R.~Akre, D.~Dowell, P.~Emma, J.~Frisch, S.~Gilevich, G.~Hays, Ph. Hering,
  R.~Iverson, C.~Limborg-Deprey, H.~Loos, A.~Miahnahri, J.~Schmerge, J.~Turner,
  J.~Welch, W.~White, and J.~Wu.
\newblock Commissioning the linac coherent light source injector.
\newblock \emph{Phys. Rev. ST Accel. Beams}, 11:\penalty0 030703, Mar 2008.
\newblock \doi{10.1103/PhysRevSTAB.11.030703}.
\newblock URL \url{https://link.aps.org/doi/10.1103/PhysRevSTAB.11.030703}.

\bibitem[Miskovich et~al.(2022{\natexlab{b}})Miskovich, Edelen, and
  Mayes]{miskovich2022}
S.~A. Miskovich, A.~Edelen, and C.~Mayes.
\newblock {PyEmittance}: A general python package for particle beam emittance
  measurements with adaptive quadrupole scans.
\newblock In \emph{The 13th International Particle Accelerator Conference}.
  IPAC22, JACoW Publishing, Geneva, Switzerland, July 2022{\natexlab{b}}.

\bibitem[Gardner et~al.(2018)Gardner, Pleiss, Weinberger, Bindel, and
  Wilson]{gardner2018gpytorch}
Jacob Gardner, Geoff Pleiss, Kilian~Q Weinberger, David Bindel, and Andrew~G
  Wilson.
\newblock Gpytorch: Blackbox matrix-matrix gaussian process inference with gpu
  acceleration.
\newblock \emph{Advances in neural information processing systems}, 31, 2018.

\bibitem[Kandasamy et~al.(2020)Kandasamy, Vysyaraju, Neiswanger, Paria,
  Collins, Schneider, Poczos, and Xing]{kandasamy2020tuning}
Kirthevasan Kandasamy, Karun~Raju Vysyaraju, Willie Neiswanger, Biswajit Paria,
  Christopher~R Collins, Jeff Schneider, Barnabas Poczos, and Eric~P Xing.
\newblock Tuning hyperparameters without grad students: Scalable and robust
  bayesian optimisation with dragonfly.
\newblock \emph{The Journal of Machine Learning Research}, 21\penalty0
  (1):\penalty0 3098--3124, 2020.

\bibitem[Hennig and Schuler(2012)]{hennig2012entropy}
Philipp Hennig and Christian~J Schuler.
\newblock Entropy search for information-efficient global optimization.
\newblock \emph{Journal of Machine Learning Research}, 13\penalty0 (6), 2012.

\bibitem[Hern{\'a}ndez-Lobato et~al.(2014)Hern{\'a}ndez-Lobato, Hoffman, and
  Ghahramani]{hernandez2014predictive}
Jos{\'e}~Miguel Hern{\'a}ndez-Lobato, Matthew~W Hoffman, and Zoubin Ghahramani.
\newblock Predictive entropy search for efficient global optimization of
  black-box functions.
\newblock \emph{Advances in neural information processing systems}, 27, 2014.

\bibitem[Wang and Jegelka(2017)]{wang2017max}
Zi~Wang and Stefanie Jegelka.
\newblock Max-value entropy search for efficient bayesian optimization.
\newblock In \emph{International Conference on Machine Learning}, pages
  3627--3635. PMLR, 2017.

\bibitem[Belakaria et~al.(2019)Belakaria, Deshwal, and Doppa]{belakaria2019max}
Syrine Belakaria, Aryan Deshwal, and Janardhan~Rao Doppa.
\newblock Max-value entropy search for multi-objective bayesian optimization.
\newblock \emph{Advances in neural information processing systems}, 32, 2019.

\bibitem[Kandasamy et~al.(2017)Kandasamy, Dasarathy, Schneider, and
  P{\'o}czos]{kandasamy2017multi}
Kirthevasan Kandasamy, Gautam Dasarathy, Jeff Schneider, and Barnab{\'a}s
  P{\'o}czos.
\newblock Multi-fidelity bayesian optimisation with continuous approximations.
\newblock In \emph{International Conference on Machine Learning}, pages
  1799--1808. PMLR, 2017.

\bibitem[Belakaria et~al.(2020)Belakaria, Deshwal, and
  Doppa]{belakaria2020multi}
Syrine Belakaria, Aryan Deshwal, and Janardhan~Rao Doppa.
\newblock Multi-fidelity multi-objective bayesian optimization: An output space
  entropy search approach.
\newblock In \emph{Proceedings of the AAAI Conference on artificial
  intelligence}, volume~34, pages 10035--10043, 2020.

\bibitem[Tomin et~al.(2016)]{Tomin:2016hnf}
Sergey Tomin et~al.
\newblock {Progress in Automatic Software-based Optimization of Accelerator
  Performance}.
\newblock In \emph{{7th International Particle Accelerator Conference}}, page
  WEPOY036, 2016.
\newblock \doi{10.18429/JACoW-IPAC2016-WEPOY036}.

\bibitem[Hanuka et~al.(2021)Hanuka, Huang, Shtalenkova, Kennedy, Edelen, Zhang,
  Lalchand, Ratner, and Duris]{hanuka2019}
Adi Hanuka, X.~Huang, J.~Shtalenkova, D.~Kennedy, A.~Edelen, Z.~Zhang, V.~R.
  Lalchand, D.~Ratner, and J.~Duris.
\newblock Physics model-informed gaussian process for online optimization of
  particle accelerators.
\newblock \emph{Phys. Rev. Accel. Beams}, 24:\penalty0 072802, Jul 2021.
\newblock \doi{10.1103/PhysRevAccelBeams.24.072802}.
\newblock URL
  \url{https://link.aps.org/doi/10.1103/PhysRevAccelBeams.24.072802}.

\bibitem[Roussel et~al.(2021{\natexlab{b}})Roussel, Gonzalez-Aguilera, Kim,
  Wisniewski, Liu, Piot, Power, Hanuka, and Edelen]{roussel2021}
Ryan Roussel, Juan~Pablo Gonzalez-Aguilera, Young-Kee Kim, Eric Wisniewski,
  Wanming Liu, Philippe Piot, John Power, Adi Hanuka, and Auralee Edelen.
\newblock Turn-key constrained parameter space exploration for particle
  accelerators using bayesian active learning.
\newblock \emph{Nature Communications}, 12\penalty0 (1):\penalty0 5612,
  September 2021{\natexlab{b}}.

\bibitem[Xu et~al.(2022)Xu, Roussel, and Edelen]{xu2022}
Connie Xu, Ryan Roussel, and Auralee Edelen.
\newblock Neural network prior mean for particle accelerator injector tuning.
\newblock In \emph{{Machine Learning and the Physical Sciences Workshop at the
  36th conference on Neural Information Processing Systems (NeurIPS)}}, 2022.
\newblock URL
  \url{https://ml4physicalsciences.github.io/2022/files/NeurIPS_ML4PS_2022_122.pdf}.

\bibitem[Newville et~al.(2019)Newville, {K Lauer}, {Dchabot}, Caswell, Gibbs,
  Péteut, Hartman, {Rokvintar}, Clarken, Martins, Allan, Birke, Jemian,
  Claesson, Adelman, Dwyer, Slepicka, Brandl, Greenberg, Vine, and {,
  André}]{pyepics}
Matt Newville, {K Lauer}, {Dchabot}, Thomas~A Caswell, Matt Gibbs, Alain
  Péteut, Steven Hartman, {Rokvintar}, Robbie Clarken, Bruno Martins, Dan
  Allan, Thomas Birke, Pete~R Jemian, Niklas Claesson, Joshua Adelman, Jack
  Dwyer, Hugo Slepicka, Georg Brandl, Eran Greenberg, David Vine, and {,
  André}.
\newblock pyepics/pyepics 3.4.0, 2019.
\newblock URL \url{https://zenodo.org/record/3241645}.

\bibitem[Loos et~al.(2010)Loos, Akre, Brachmann, Coffee, Decker, Ding, Dowell,
  Edstrom, Emma, Fisher, Frisch, Gilevich, Hays, Hering, Huang, Iverson,
  Messerschmidt, Miahnahri, Moeller, Nuhn, Ratner, and SLAC]{osti_982082}
Henrik Loos, R.~Akre, A.~Brachmann, R.~Coffee, F.~J. Decker, Y.~Ding,
  D.~Dowell, S.~Edstrom, P.~Emma, A.~Fisher, J.~Frisch, S.~Gilevich, G.~Hays,
  Ph. Hering, Z.~Huang, R.~Iverson, M.~Messerschmidt, A.~Miahnahri, S.~Moeller,
  H.~D. Nuhn, D.~Ratner, and Livermore SLAC, LLNL.
\newblock Operational performance of lcls beam instrumentation.
\newblock 6 2010.
\newblock URL \url{https://www.osti.gov/biblio/982082}.

\bibitem[Loehl(2005)]{Loehl2005-xc}
F.~Loehl.
\newblock Measurements of the transverse emittance at the {VUV-FEL}.
\newblock Technical Report 1435-8085, Germany, 2005.

\bibitem[Xie(1996)]{xie1995}
M.~Xie.
\newblock {Design optimization for an X-ray free electron laser driven by SLAC
  LINAC}.
\newblock \emph{Conf. Proc. C}, 950501:\penalty0 183--185, 1996.
\newblock \doi{10.1109/PAC.1995.504603}.

\bibitem[Ratner et~al.(2010)Ratner, Brachmann, Decker, Ding, Dowell, Emma,
  Frisch, Gilevich, Hays, Hering, Huang, Iverson, Loos, Miahnahri, Nuhn,
  Turner, Welch, White, Wu, Xiang, and Yocky]{Ratner2010}
Daniel Ratner, A.~Brachmann, F.J. Decker, Y.~Ding, D.~Dowell, P.~Emma,
  J.~Frisch, S.~Gilevich, G.~Hays, P.~Hering, Z.~Huang, R.~Iverson, H.~Loos,
  A.~Miahnahri, H.D. Nuhn, J.~Turner, J.~Welch, W.~White, J.~Wu, D.~Xiang, and
  G.~Yocky.
\newblock {FEL Gain Length and Taper Measurements at LCLS}.
\newblock In \emph{{31st International Free Electron Laser Conference,
  TUOA03}}, 7 2010.
\newblock URL \url{https://epaper.kek.jp/FEL2009/papers/tuoa03.pdf}.

\bibitem[Moosbauer et~al.(2022)Moosbauer, Casalicchio, Lindauer, and
  Bischl]{moosbauer2022enhancing}
Julia Moosbauer, Giuseppe Casalicchio, Marius Lindauer, and Bernd Bischl.
\newblock Enhancing explainability of hyperparameter optimization via bayesian
  algorithm execution.
\newblock \emph{arXiv preprint arXiv:2206.05447}, 2022.

\bibitem[Mehta et~al.(2021)Mehta, Paria, Schneider, Ermon, and
  Neiswanger]{mehta2021experimental}
Viraj Mehta, Biswajit Paria, Jeff Schneider, Stefano Ermon, and Willie
  Neiswanger.
\newblock An experimental design perspective on model-based reinforcement
  learning.
\newblock In \emph{International Conference on Learning Representations}, 2021.

\bibitem[Kandasamy et~al.(2019)Kandasamy, Neiswanger, Zhang, Krishnamurthy,
  Schneider, and Poczos]{kandasamy2019myopic}
Kirthevasan Kandasamy, Willie Neiswanger, Reed Zhang, Akshay Krishnamurthy,
  Jeff Schneider, and Barnabas Poczos.
\newblock Myopic posterior sampling for adaptive goal oriented design of
  experiments.
\newblock In \emph{International Conference on Machine Learning}, pages
  3222--3232. PMLR, 2019.

\bibitem[Qiang et~al.(2006)Qiang, Lidia, Ryne, and Limborg-Deprey]{impact}
Ji~Qiang, Steve Lidia, Robert~D. Ryne, and Cecile Limborg-Deprey.
\newblock Three-dimensional quasistatic model for high brightness beam dynamics
  simulation.
\newblock \emph{Phys. Rev. ST Accel. Beams}, 9:\penalty0 044204, Apr 2006.
\newblock \doi{10.1103/PhysRevSTAB.9.044204}.
\newblock URL \url{https://link.aps.org/doi/10.1103/PhysRevSTAB.9.044204}.

\bibitem[Nogueira(2014)]{bayesopt}
Fernando Nogueira.
\newblock {Bayesian Optimization}: Open source constrained global optimization
  tool for {Python} (version 1.1), 2014.
\newblock URL \url{https://github.com/fmfn/BayesianOptimization}.

\bibitem[Virtanen et~al.(2020)Virtanen, Gommers, Oliphant, Haberland, Reddy,
  Cournapeau, Burovski, Peterson, Weckesser, Bright, {van der Walt}, Brett,
  Wilson, Millman, Mayorov, Nelson, Jones, Kern, Larson, Carey, Polat, Feng,
  Moore, {VanderPlas}, Laxalde, Perktold, Cimrman, Henriksen, Quintero, Harris,
  Archibald, Ribeiro, Pedregosa, {van Mulbregt}, and {SciPy 1.0
  Contributors}]{2020SciPy-NMeth}
Pauli Virtanen, Ralf Gommers, Travis~E. Oliphant, Matt Haberland, Tyler Reddy,
  David Cournapeau, Evgeni Burovski, Pearu Peterson, Warren Weckesser, Jonathan
  Bright, St{\'e}fan~J. {van der Walt}, Matthew Brett, Joshua Wilson, K.~Jarrod
  Millman, Nikolay Mayorov, Andrew R.~J. Nelson, Eric Jones, Robert Kern, Eric
  Larson, C~J Carey, {\.I}lhan Polat, Yu~Feng, Eric~W. Moore, Jake
  {VanderPlas}, Denis Laxalde, Josef Perktold, Robert Cimrman, Ian Henriksen,
  E.~A. Quintero, Charles~R. Harris, Anne~M. Archibald, Ant{\^o}nio~H. Ribeiro,
  Fabian Pedregosa, Paul {van Mulbregt}, and {SciPy 1.0 Contributors}.
\newblock {{SciPy} 1.0: Fundamental Algorithms for Scientific Computing in
  Python}, (version 1.7.3).
\newblock \emph{Nature Methods}, 17:\penalty0 261--272, 2020.
\newblock \doi{10.1038/s41592-019-0686-2}.

\end{thebibliography}

\appendix

\section{Experimental Setup}\label{appendixexpsetup}

\textit{LCLS Injector Overview.} The LCLS injector begins with a photocathode RF gun followed by two S-band (2.856 GHz) linac sections (L0a and L0b) producing one or more electron bunches at a 120 Hz repetition rate with an energy of 135 MeV (see Figure~\ref{fig:layout}) \cite{lclsinj}. In this work, the electron bunch charge was 250 pC. The algorithm and analysis software communicated with experimental devices using PyEpics \cite{pyepics}, a Python interface to the EPICS Channel Access (CA) library for the EPICS control system.

\textit{LCLS RF Gun and Measurement Quadrupole.} The focusing gun solenoid (SOL1) is located directly downstream of the RF gun and upstream of L0a, and has two small quadrupole coils incorporated into its magnet. These coils, one normal (CQ1) and one skew (SQ1) quadrupole, are single wires spanning the length of the solenoid, and are intended to cancel a small quadrupole field error at the end of SOL1. The measurement (scanning) quadrupole (Q5) is 11.7 m from the gun solenoid, and is separated from the wire scanner by a 2.1-m long drift. Figure \ref{fig:layout} shows an illustration of this layout. 

\textit{LCLS Beam Diagnostics.} 
The transverse beam profiles at LCLS were measured with minimally-invasive wire scanners for both $X$- and $Y$-planes. They require multiple shots for each measurement, and provide a multi-pulse averaged integrated beam profile from photomultiplier tube measurements where a Gaussian distribution was used to extract the RMS beam size from the integrated profiles. The statistical fitting error of 3\% on average was propagated to the emittance result. The systematic error from the choice of beam size distribution, not taken into account in the experimental emittance results, was estimated to be around 4\%/7\% in $X$-/$Y$-plane. Each beam-size measurement took at least 18 seconds to execute a wire scanner measurement. While optical transition radiation (OTR) screens would be faster than wire scanners, only the slow wire measurements were available at LCLS at the time of this work.

\textit{FACET-II Setup.}
The FACET-II injector produces a single electron bunch at a 120 Hz repetition rate with an energy of 125 MeV. In this setup, the measurement quadrupole is 2.7 m from the measurement screen, separated only by a drift. The beam profiles were measured with an OTR screen. For each measurement at each quadrupole setting, after a 3 second wait-time for the magnet to settle, four images were acquired, background subtracted, then averaged. The $X$- and $Y$-plane beam profile projections of the final image were fitted with a Gaussian distribution whose width was taken as the beam size. The statistical error from the fit (around 3\% on average) was propagated to the emittance result. 

\section{Beam Emittance Calculations}\label{appendixemitmeas}
In this work, we only consider the transverse emittance in the $X$- and $Y$-dimensions, which represents the area of the beam in the transverse planes defined by the positions and momenta, $x$ and $p_x$ in the horizontal plane, and $y$ and $p_y$ in the vertical plane (the $X$- and $Y$-axes are defined as orthogonal and the $Z$-axis as parallel to the path of the beam). All mentions of emittance in this article refer to the geometric mean of the normalized transverse emittance, unless otherwise noted. 

Common methods to measure electron-beam size in the transverse planes involve either scanning a wire through the beam or passing the beam through a yttrium aluminum garnet (YAG) or OTR screen \cite{osti_982082}. An emittance calculation is done using beam-size measurements with either multiple wires/screens at different positions in the accelerator, or, as we used in this work, a single wire/screen while changing an upstream quadrupole magnet \cite{minty,Loehl2005-xc}. Here, each full emittance calculation was performed using the Q5 quadrupole with adaptive scanning methods as outlined in \cite{miskovich2022}. For each full emittance calculation executed on the machine, four initial points are measured as a rough scan along the full quadrupole range to find the approximate waist location, which can change as the upstream injector configuration is varied. Once the waist is found, a finer scan of seven points is measured around the waist, with redundant points skipped during acquisition. During analysis, any data outside of the convex region are removed from the set based on the inflection points, and points are added as needed to recover a symmetric parabola. The process repeats in each of the $X$ and $Y$ dimensions, for a total of 18 beam-size measurements on average.

When accelerator operators tune the LCLS photoinjector by hand, emittance calculations typically involve a scan of at least four measurements over the domain of the scanning quadrupole upstream of the wire. However, in some configurations of the injector, the waist of the beam--a necessary feature to capture--in the transverse $X$- and $Y$-plane does not occur at the same quadrupole strength, thus requiring two quadrupole scans over two different ranges (or one large range). Moreover, to incorporate the calculation into automated tuning, the calculation must be robust, requiring additional measurements. To achieve robust and accurate measurements, the quadrupole scan range needs to be large enough to fully capture the waist of the beam along both axes, while still being narrow enough to accurately quantify the minimum of the parabolic curve. This range can vary for different machine settings and is typically adjusted by hand until a good scan is achieved.

Here, the number of measurements per scan executed on the machine (18 on average) was empirically chosen. Using fewer beam-size measurements per scan was insufficiently robust during optimization, leading to erroneous emittance calculations. Using additional measurements per scan resulted in a slightly smaller final emittance, but at the cost of even slower convergence. BAX uses 30 measurements to calculate the virtual objective on the internal model since acquisition time is many orders of magnitude faster than querying the machine.

\begin{figure}
\centering
\includegraphics[width=0.55\linewidth]{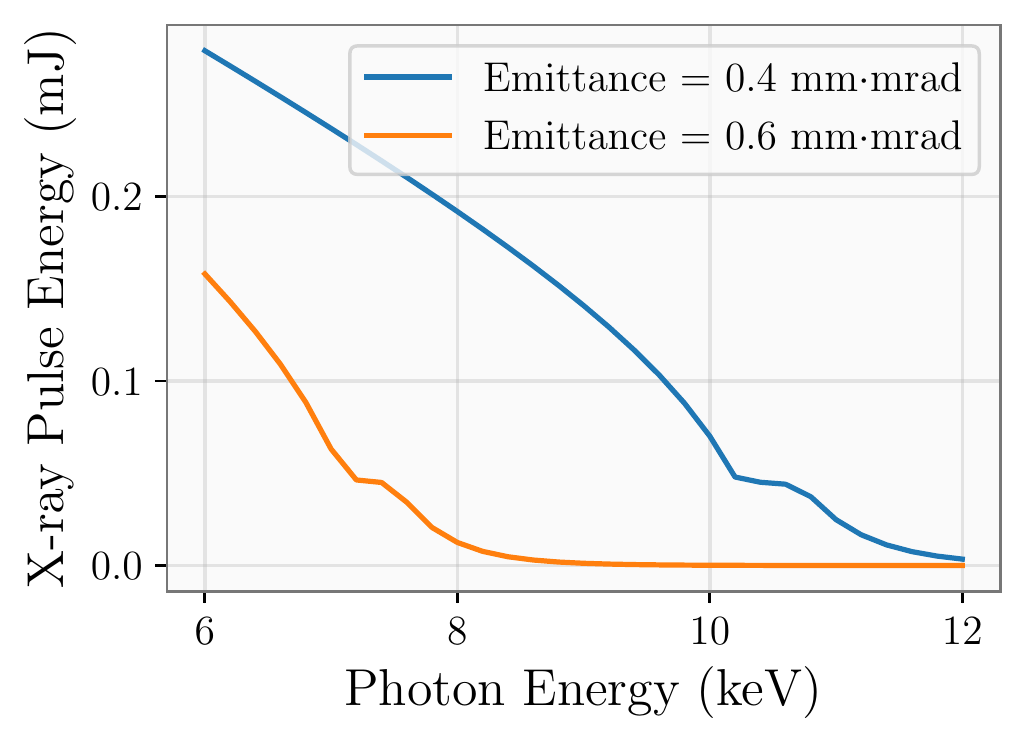}
\caption{\label{fig:lclsii-estimate} LCLS-II-HE HXR performance estimate for two different emittance values, using a Ming-Xie model \cite{xie1995} with an electron beam energy of
8 GeV, a bunch charge of 50 pC, a peak current of 500 A, an integrated gradient of the focusing-defocusing (FODO) quadrupole of 3 T, and an undulator period length of 26 mm.}
\end{figure}

\section{LCLS-II-HE Emittance Calculations}
\label{appendixlclsii}
Figure~\ref{fig:lclsii-estimate} shows calculations of X-ray FEL pulse intensities at the end of the hard X-ray (HXR) undulator line at LCLS-II high-energy upgrade (LCLS-II-HE) .
The estimates use a model based on the Ming Xie FEL function \cite{xie1995, Ratner2010} that was first developed to match the measured LCLS beam performance. The same model has also been applied to the soft X-ray (SXR) and HXR LCLS-II undulator lines running with beams accelerated by the copper (normal-conducting) linac, and again confirmed to match the measured beam performance measurements. In Figure~\ref{fig:lclsii-estimate}, the model is applied to calculate the final FEL power as a function of photon energy from a beam accelerated by the future superconducting LCLS-II-HE linac.  Even a moderate increase in emittance from 0.4 to 0.6 mm$\cdot$mrad has a significant effect on LCLS-II-HE performance, because the FEL gain length increases with higher emittance as well as higher photon energy. At a critical photon energy, the undulator line is too short for the FEL to saturate, and the power drops dramatically for higher photon energies. In the case shown in the figure, decreasing the slice emittance from 0.6 mm$\cdot$mrad to 0.4 mm$\cdot$mrad increases the maximum photon energy at which the FEL can reach saturation from 7.3 keV to 10.5 keV, high enough for atomic resolution in single particle imaging.
The lower emittance also results in a significant increase in X-ray pulse intensity at all photon energies.
While LCLS-II-HE is still in the design phase and model accuracy is uncertain, the success with LCLS and LCLS-II gives confidence to the Ming Xie calculations.

\section{\textsc{Multipoint-BAX} and its Relation to BAX Algorithms}\label{appendixbax}
We provide this section to give additional context for readers unfamiliar with the method of Bayesian algorithm execution (BAX) \citep{neiswanger2021bayesian}, and the InfoBAX acquisition function.

We can view our multipoint black-box optimization setting---which is a joint measurement and optimization problem---as the task of inferring a certain \textit{computable property $\mathcal{O_A}$} of a potentially noisy black-box function $f$, produced by running an algorithm $\mathcal{A}$ on $f$. In our case, the computable property $\mathcal{O_A}$ is the minimizer of the emittance (and the corresponding algorithm $\mathcal{A}$ is an emittance minimization algorithm), while the black box function $f$ is the beam size function, which maps each point in the joint control-measurement domain to a given beam size.

The method of Bayesian algorithm execution (BAX) \citep{neiswanger2021bayesian} presents a general framework for inferring computable properties $\mathcal{O_A}$ within $\mathcal{T}$ function evaluations, given $\mathcal{A}$ and a prior distribution on $f$, $p(f)$, that captures the initial uncertainty about the true function. In this work, $p(f)$ is defined by a Gaussian process (GP) model, and we denote $p(f|\mathcal{D}_t)$ as the posterior distribution of $f$ given a dataset $\mathcal{D}_t$. This dataset $\mathcal{D}_t$ consists of any initial observations along with $t-1$ queried observations. We then use $p(\mathcal{O_A}|\mathcal{D}_t)$ to denote the induced posterior distribution over the algorithm output $\mathcal{O_A}$.

In order to reduce the number of function queries, our \textsc{Multipoint-BAX} method uses techniques from an information-based BAX method, InfoBAX \cite{neiswanger2021bayesian, moosbauer2022enhancing, mehta2021experimental}, to make targeted queries that maximize the mutual-information between $\mathcal{O_A}$ and the next observation $y_{t}$. The InfoBAX procedure is a sequential algorithm that seeks to maximize the acquisition function, defined here as the expected information gain (EIG) about $\mathcal{O_A}$ upon observing $y_{t}$. The next sampling point is then chosen to be the $x_t$ that maximizes the estimated EIG. We can write EIG as
\begin{equation*}
    \text{EIG}_t (x) = \text{H}[\mathcal{O_A}|\mathcal{D}_t]-\mathbb{E}_{p(y_x|\mathcal{D}_t)} [\text{H}[\mathcal{O_A}|\mathcal{D}_t \cup \{(x,y_x)\}]],
\end{equation*}
where $\text{H}[\mathcal{O_A}|\mathcal{D}_t]$ is the (Shannon) entropy of $p(\mathcal{O_A}|\mathcal{D}_t)$, and $p(y_x|\mathcal{D}_t)$ is the posterior predictive distribution at $x$ given $\mathcal{D}_t$. For details on how EIG$(x)$ is estimated, we refer the reader to \cite{neiswanger2021bayesian}. 

To compute EIG efficiently, the InfoBAX procedure only executes the algorithm $\mathcal{A}$ on function samples $\tilde f$ drawn from the posterior distribution $p(f|\mathcal{D}_t)$, similar to other posterior sampling-based methods for experimental design \cite{hernandez2014predictive, kandasamy2019myopic}. In this manner, the true algorithm output is inferred with minimal evaluations of the true (expensive) function $f$. Once the next sampling point is selected at each iteration, a single evaluation of the true function is made at that selected point.

\section{Computational Considerations}\label{appendixtiming}
One potential hurdle for the implementation of the \textsc{Multipoint-BAX} algorithm is the computational complexity, as each iteration requires repeated calculation of the virtual objective on draws from the model posterior, which in our case can require a large number of emittance calculations. Although this work did not focus on computational efficiency of the algorithm, in cases where the virtual objective is expensive, parallelization of the \textsc{Multipoint-BAX} algorithm execution procedure can significantly reduce wall-clock time. During live optimization at LCLS, a parallelized version using five CPU cores takes 4.7 seconds, on average, to select a beam-size function query at each iteration of the algorithm.

\section{LCLS Injector Surrogate Model}\label{appendixinjsurr}
The LCLS injector surrogate model used in this work is a neural network (NN) based surrogate model that provides fast, non-invasive predictions of electron-beam properties. The NN was trained on IMPACT-T simulation \cite{impact} data using 16 input parameters including the pulse length, laser radius, gun solenoid and quadrupole settings, L0 linac phase, and all matching quadrupoles settings. The NN architecture consists of 17 layers, and the model outputs scalar predictions of the $X$- and $Y$-plane beam sizes. To mimic experimental emittance calculation procedures in a simulation setting, the $X$-, $Y$-plane beam sizes were used when performing a beam size measurement, and Gaussian noise $\mathcal{N}(0,\sigma_n)$ was added to each beam size before passing the measurements to the emittance calculation.

\section{Details of Algorithm Comparison in Simulation Setting}\label{appendixalgo}
Bayesian optimization was implemented using the \verb|Bayesian Optimization| package \cite{bayesopt}. We defined the BO's surrogate model as a GP constructed with an RBF kernel, and the upper confidence bound (UCB) as the acquisition function with an exploration weight of 2.0 which was maximized to select the next function query. For each run, BO was initialized with three random scans that were uniformly and randomly sampled from the control domain. The simplex algorithm was implemented using the optimization routine available through \verb|SciPy| \cite{2020SciPy-NMeth}. Each simplex optimization was run with an initial guess of the injector configuration that was randomly sampled from the control domain as well.

In some cases when using the NN and querying points far from optimal, the emittance calculation fails to return a number. Such cases were handled by assigning a large number to simplex queries, and the \textsc{Multipoint-BAX} and BO algorithms were set to pass any failed measurements and move to the next optimal point in the acquisition function optimization. Additionally, any emittance calculations with an uncertainty larger than 70\% were ignored by the BO and \textsc{Multipoint-BAX} optimizers.

\end{document}